\documentclass{cls/IEEEojcsys}

\usepackage[table,usenames,dvipsnames]{xcolor}      
\usepackage[noadjust]{cite}
\usepackage{amsmath,amssymb,amsfonts,amsthm,dsfont,mathtools}

\usepackage{graphicx,tabularx,adjustbox}
\usepackage{multirow}
\usepackage[font=small]{caption}
\usepackage[font=footnotesize]{subcaption}

\usepackage{algorithmic}
\usepackage[ruled,vlined]{algorithm2e}

\usepackage{dirtytalk}
\allowdisplaybreaks

\usepackage[breaklinks=true, colorlinks, bookmarks=true, citecolor=Black, urlcolor=Violet,linkcolor=Black]{hyperref}

\newcommand{\new}[1]{\textcolor{blue}{#1}}

\newtheorem{theorem}{Theorem}[section]
\newtheorem{proposition}[theorem]{Proposition}
\newtheorem{lemma}[theorem]{Lemma}

\newtheorem{remark}[theorem]{Remark}
\newtheorem{definition}[theorem]{Definition}
\newtheorem{problem}{Problem}
\newtheorem*{problem*}{Problem}
\newtheorem{assumption}{Assumption}

\newcommand{\norm}[1]{\lVert#1\rVert}
\newcommand\xqed[1]{%
  \leavevmode\unskip\penalty9999 \hbox{}\nobreak\hfill
  \quad\hbox{#1}}
\newcommand\demo{\xqed{$\bullet$}}

\newcommand{\calA}{\mathcal{A}}
\newcommand{\calB}{\mathcal{B}}

\newcommand{\calD}{\mathcal{D}}

\newcommand{\calF}{\mathcal{F}}

\newcommand{\calM}{\mathcal{M}}
\newcommand{\calN}{\mathcal{N}}

\newcommand{\calP}{\mathcal{P}}

\newcommand{\calU}{\mathcal{U}}

\newcommand{\calX}{\mathcal{X}}

\newcommand{\bff}{\mathbf{f}}

\newcommand{\bfu}{\mathbf{u}}

\newcommand{\bfw}{\mathbf{w}}
\newcommand{\bfx}{\mathbf{x}}
\newcommand{\bfy}{\mathbf{y}}

\newcommand{\bftheta}{\boldsymbol{\theta}}

\newcommand{\bfpi}{\boldsymbol{\pi}}

\newcommand{\bfxi}{\boldsymbol{\xi}}

\newcommand{\bfI}{\mathbf{I}}

\newcommand{\bfW}{\mathbf{W}}
\newcommand{\bfX}{\mathbf{X}}

\newcommand{\bbE}{\mathbb{E}}

\newcommand{\bbN}{\mathbb{N}}

\newcommand{\bbP}{\mathbb{P}}
\newcommand{\bbQ}{\mathbb{Q}}
\newcommand{\bbR}{\mathbb{R}}

\begin{document}
\sptitle{Article Category}

\title{Distributionally Robust Policy and Lyapunov-Certificate Learning}

\author{Kehan Long\affilmark{1} (Student Member)}

\author{Jorge Cort{\'e}s\affilmark{1} (Fellow)}

\author{Nikolay Atanasov\affilmark{1} (Senior Member)}

\affil{Contextual Robotics Institute, University of California San Diego, La Jolla, CA 92093, USA} 

\corresp{CORRESPONDING AUTHOR: Kehan Long (e-mail: \href{mailto:k3long@ucsd.edu}{k3long@ucsd.edu})}
\authornote{This work was supported by ONR Award N00014-23-1-2353 and NSF CCF-2112665 (TILOS).}

\markboth{Distributionally Robust Policy and Lyapunov-Certificate Learning}{LONG {\itshape ET AL}.}

\begin{abstract}
This article presents novel methods for synthesizing distributionally robust stabilizing neural controllers and certificates for control systems under model uncertainty. A key challenge in designing controllers with stability guarantees for uncertain systems is the accurate determination of and adaptation to shifts in model parametric uncertainty during online deployment. We tackle this with a novel distributionally robust formulation of the Lyapunov derivative chance constraint ensuring a monotonic decrease of the Lyapunov certificate. To avoid the computational complexity involved in dealing with the space of probability measures, we identify a sufficient condition in the form of deterministic convex constraints that ensures the Lyapunov derivative constraint is satisfied. We integrate this condition into a loss function for training a neural network-based controller and show that, for the resulting closed-loop system, the global asymptotic stability of its equilibrium can be certified with high confidence, even with Out-of-Distribution (OoD) model uncertainties. To demonstrate the efficacy and efficiency of the proposed methodology, we compare it with an uncertainty-agnostic baseline approach and several reinforcement learning approaches in two control problems in simulation. Open-source implementations of the examples are available at \href{https://github.com/KehanLong/DR_Stabilizing_Policy}{https://github.com/KehanLong/DR\_Stabilizing\_Policy}.
\end{abstract}

\begin{IEEEkeywords}
Learning for control, Lyapunov methods, optimization under uncertainty, stability of nonlinear systems
\end{IEEEkeywords}

\maketitle

\renewcommand{\thefootnote}{\fnsymbol{footnote}}
\footnotetext{A preliminary version of this paper appeared as~\cite{long2023dro_lf} at the 2023 Learning for Dynamics and Control Conference.}


%
\section{Introduction}\label{sec: intro}

In control theory and robotics, the task of synthesizing stabilizing controllers with provable certificates for open-loop systems under model uncertainty remains an important challenge. A Lyapunov function (LF) \cite{slotine1991applied} is a fundamental tool for asserting the asymptotic stability of ordinary differential equations (ODEs). 
Existing approaches in deriving stabilizing controllers for uncertain nonlinear systems usually rely on known bounds or probabilistic models of the uncertainty, which may be hard to obtain in practice. In this work, we address this challenge by exploring the potential of neural networks to synthesize distributionally robust stabilizing controllers and associated Lyapnuov stability certificates, providing a new approach for ensuring robustness to model uncertainty. 

The use of Lyapunov functions for ensuring the stability of nonlinear systems is well documented in the literature \cite{Artstein1983StabilizationWR, SONTAG1989117, haddad2008nonlinear}. The early 2000s marked a significant development in the field with the introduction of sum-of-square (SOS) polynomials to identify LFs through semi-definite programming (SDP) \cite{parrilo2000structured, Papachristo_2002_sos_lf}. These approaches, however, often necessitate polynomial approximations of the system dynamics and scale poorly in higher dimensional spaces. Moreover, traditional SOS methods face difficulties in the joint synthesis of stabilizers and LF certificates for open-loop systems \cite{jarvis_clf_sos}. With the remarkable advancements in using neural networks as function approximators, researchers \cite{Chang2019NeuralLC, Richards2018TheLN, dai_2021_lyapunov, gaby2021lyapunov_net, li2023task, wu2023neural, zhang2024learning} have started representing LFs and/or control policies using neural networks. This progress holds great potential for developing stabilizing controllers and certificates for complex nonlinear systems, offering a vastly expanded functional space compared to SOS polynomials. Despite these advancements, a critical gap persists in deriving controllers equipped with suitable stability guarantees for systems with model uncertainty, an aspect paramount to the practical application of control and robotic systems in real-world scenarios. To address the challenges of ensuring stability and stabilization in systems with model uncertainty, a related class of works \cite{Taylor_2019iros, Castaneda_GPCLF_ACC21, Long2022RAL, PM-JC:23-auto, PM-KL-NA-JC:23-csl, Long_learningcbf_ral21} has focused on adapting stability certificates for nominal systems by incorporating model uncertainty during deployment. These approaches commonly assume knowing the nominal certificates or controllers as well as the error bounds or distribution of the model uncertainty. 


Our work aims to develop a distributionally robust Lyapunov function and controller pair for systems with model uncertainty. Instead of relying on knowledge of the distribution or error bounds of the model uncertainty, we seek to synthesize a controller that ensures the global asymptotic stability of the controlled system, even in the presence of distribution shifts during online deployment. To do this, we rely on recent advances in distributionally robust optimization (DRO) \cite{AS-DD-AR:21, Esfahani2018DatadrivenDR}. Distributionally robust chance constraints handle uncertainty in constraints with a limited number of available samples. The core strategy involves constructing an ambiguity set around the empirical distribution derived from these samples, defined by a radius using a probability distance measure such as the Wasserstein distance \cite{Esfahani2018DatadrivenDR, Hota2019DataDrivenCC}. The objective is to ensure constraint satisfaction with high confidence for all distributions within this set. This key benefit of the approach is its robust performance guarantee for distributional uncertainty, i.e., uncertainty in the possible probability distribution of the samples. 

The growing field of distributionally robust control (DRC) \cite{yang2020wasserstein,coulson2021distributionally, bahari_2022_safedro_rl,lathrop2021distributionally,Hakobyan_dr_ddp, li2021distributionally,aolaritei2023wasserstein,li2023_drc_output, micheli2022data, taskesen2024distributionally} offers tools for control design under uncertainty. DRC has been successfully applied to various control domains, including model predictive control \cite{aolaritei2023wasserstein, micheli2022data, li2023_drc_output}, differential dynamic programming \cite{Hakobyan_dr_ddp}, and linear quadratic control \cite{taskesen2024distributionally}. The main objective of DRC is to design controllers that provide robust performance and constraint satisfaction under distributional shifts, making them well-suited for real-world applications where the true uncertainty distribution may differ from the estimated distribution. Building upon the principles of DRC, our work focuses on developing a distributionally robust LF and controller pair for control systems with model uncertainty.

%
%

This paper is an extended version of our previous conference publication~\cite{long2023dro_lf}, which introduced distributionally robust Lyapunov stability for closed-loop systems with model uncertainty. In this work, we extend our approach to the joint synthesis of distributionally robust controllers and Lyapunov certificates for nonlinear control systems. Our \textbf{contributions} include the following.
\begin{itemize}
    \item Given samples of uncertain model parameters, we propose a method for joint synthesis of a distributionally robust controller and a Lyapunov certificate for a nonlinear control system subject to affine model uncertainty;

    \item We identify a sufficient convex constraint that ensures the distributionally robust Lyapunov condition is satisfied and employ it to learn a stabilizing neural network controller and Lyapunov function pair;

    \item We show that the resulting neural network controller guarantees the global asymptotic stability in probability of the closed-loop system;

    \item We validate in both theory and experiments that uncertain systems governed by the learned controller are stable with high confidence, even under out-of-distribution model uncertainty.  
\end{itemize}


\section{Related Work}
\label{sec: related_work}


This section reviews related work on synthesizing Lyapunov-stable controllers and distributionally robust optimization. 


\textbf{Stabilizing Control Design.}
Lyapunov theory plays a pivotal role in synthesizing stabilizing controllers for nonlinear systems. It offers strong asymptotic stability guarantees for the controlled system, which are particularly useful in real-world applications. For linear systems, synthesizing Lyapunov-stable controllers can be efficiently accomplished by linear quadratic regulator techniques \cite{Boyd_LMI_control}. For control-affine systems with polynomial dynamics, researchers \cite{jarvis_clf_sos} have successfully obtained controller and certificate pairs by solving SOS programs. However, the absence of a valid SOS Lyapunov function does not necessarily imply the system's instability, as certain positive-definite functions are not SOS-representable \cite{HilbertUeberDD}. Moreover, the computational complexity of SOS techniques limits their applicability to relatively simple polynomial systems.

Recently, there has been an increasing focus on leveraging neural networks to address the limitations of SOS-based methods. Neural networks offer more general function representations compared to SOS polynomials, thereby enhancing the capacity to handle a broader class of systems.

Richards \textit{et al}. \cite{Richards2018TheLN} introduced a neural network-based method to learn the region of attraction for given controllers in discrete-time systems. Chang \textit{et al}. \cite{Chang2019NeuralLC} focused on synthesizing controllers and neural network Lyapunov functions for nonlinear continuous-time systems, with verification and improvement of the learned functions by satisfiability modulo theories (SMT) solvers. Boffi \textit{et al}. \cite{boffi2021learning} improved the efficiency of learning Lyapunov functions by integrating positive-definiteness and equilibrium conditions directly into the network architecture. Dai \textit{et al}. \cite{dai_2021_lyapunov} developed an approach for discrete-time systems that synthesizes neural network stabilizing controllers and Lyapunov functions, refining controller performance using mixed integer programs (MIP) verifiers. Gaby \textit{et al}. \cite{gaby2021lyapunov_net} theoretically certified neural network Lyapunov functions by analyzing the approximation power of networks with specific architectures. Zhou \textit{et al}. \cite{zhou2022neural} proposed a framework for learning nonlinear systems, stabilizing controllers, and Lyapunov functions with stability guarantees verified by SMT solvers. Dawson \textit{et al}. \cite{dawson_2022_robust_CLBF} extended these ideas to learning safety certificates as control Lyapunov-barrier functions for control-affine systems with bounded convex-hull uncertainty. A more extensive survey by Dawson \textit{et al}. \cite{Dawson_2022_survey} details recent advancements in this line of research. Another related line of research is the use of CLFs to improve the sample efficiency and robustness of reinforcement learning (RL) algorithms \cite{westenbroek2022lyapunov, lopez2024decomposing}.

Despite these developments, there has been limited focus on synthesizing Lyapunov-stable controllers with provable certificates for systems under model uncertainty, particularly in settings where bounds or the distribution of the model uncertainty are challenging to determine.

\textbf{Distributionally Robust Optimization.}
Distributionally robust optimization has emerged as a powerful approach to tackle uncertainty in optimization problems, particularly advantageous when dealing with a limited number of samples. To account for the discrepancy between the empirical and true distributions of the uncertainty samples, researchers have introduced several uncertainty descriptors, such as moment ambiguity sets~\cite{Parys2015monent}, Kullback–Leibler ambiguity sets~\cite{Jiang2016DatadrivenCC}, and Wasserstein ambiguity sets~\cite{Esfahani2018DatadrivenDR, Xie2021OnDR, Hota2019DataDrivenCC}. The DRO framework has been increasingly recognized for its robust performance guarantees against distributional uncertainty, leading to its widespread application in machine learning~\cite{sagawa2019distributionally, levine2020offline}, uncertainty quantification~\cite{FB-DB-JC-SM-DMT:21}, control theory~\cite{DB-JC-SM:21-tac,li2021distributionally,ashish_2023_dro, yang2020wasserstein,DB-JC-SM:24-tac}, and robotics~\cite{ren2022distributionally_ral, coulson2021distributionally}. 

Long \textit{et al}.~\cite{long2023_clf_cbf_drccp} proposed a distributionally robust formulation for safe stabilizing control under model uncertainty, assuming nominal safe and stable certificates are provided. Ren and Majumdar \cite{ren2022distributionally_ral} introduced a DRO framework that enhances policy robustness by iteratively training with adversarial environments generated through a learned generative model. Liviu \textit{et al}. \cite{aolaritei2023wasserstein} introduced Wasserstein tube MPC for stochastic systems, enhancing robustness and efficiency by effectively analyzing state trajectory uncertainties with limited noise samples. In~\cite{bahari_2022_safedro_rl}, a distributionally robust model predictive control formulation with the Wasserstein metric is proposed, with its parameters efficiently optimized through reinforcement learning. Hakobyan and Yang~\cite{Hakobyan_dr_ddp} developed a tractable distributionally robust differential dynamical programming method for stochastic systems by utilizing the Kantrovich duality principle. Distributionally robust RRT for 
motion planning that addresses model and environmental uncertainties was developed using moment-based ambiguity sets by Summers \cite{summers2018_dr_rrt} and using Wasserstein-metric-based ambiguity sets by Lathrop \textit{et al} \cite{lathrop2021distributionally}.



\section{Background}\label{sec: prelim}

This section introduces our notation and offers a brief review of Lyapunov functions.

\subsection{Notation}
We denote the sets of real, non-negative real, and natural numbers by $\bbR$, $\bbR_{\geq 0}$, and $\bbN$, respectively. For $N \in \bbN$, we let $[N] := \{1,2, \dots N\}$. For a scalar $x$, we define $(x)_+ := \max(x,0)$. We denote by $\bfI_n \in \bbR^{n \times n}$ the identity matrix and $\boldsymbol{0}_n \in \bbR^n$ the zero vector. For a vector $\bfx$ and a matrix $\bfX$, we use $\|\bfx\|$ and $\|\bfX\|$ to denote the $L_2$ norm and the spectral norm, respectively.
We use $B(\bfx; \delta)$ to denote the open ball centered at $\bfx$ with radius $\delta$. The closure of $B(\bfx; \delta)$ is denoted by $\overline{B}(\bfx; \delta)$.
%
%
The gradient of a differentiable function $V$ is denoted by $\nabla V$.
We consider a complete separable metric space $\Xi$ with metric $d$ and Borel $\sigma$-algebra $\calF$. The set of Borel probability measures on $\Xi$ is denoted as $\calP(\Xi)$. For a random variable $\bfxi$ with distribution supported on $\Xi$, we denote the distribution and expectation by $\mathbb{P}^*$ and $\bbE_{\bbP^*}(\bfxi)$, respectively.

\subsection{System Stabilization Certified by Lyapunov Functions}
\label{sec: lyapunov_original}
Consider the controlled continuous-time system
\begin{equation}
\label{eq: open_loop_control_system}
\dot{\bfx} = \bff(\bfx, \bfu),
\end{equation}
where $\bfx \in \calX \subset \bbR^n$ denotes the system state and $\bfu \in \mathcal{U} \subset \mathbb{R}^{m}$ is the control input. We assume $\bff: \mathbb{R}^{n} \times \bbR^m \mapsto \mathbb{R}^{n}$ is locally Lipschitz, and the origin $\bfx = \mathbf{0}_n$ is the desired equilibrium of the unforced system, i.e., $f(\boldsymbol{0}_n, \boldsymbol{0}_m) = \boldsymbol{0}_n$.
Plugging a Lipschitz controller $\bfu = \bfpi(\bfx)$ in \new{\eqref{eq: open_loop_control_system}} results in the closed-loop system
\begin{equation}
\label{eq: closed_loop_system}
    \dot{\bfx} = \bff_{\text{cl}}(\bfx) := \bff(\bfx, \bfpi(\bfx)). 
\end{equation}
In this context, a Lyapunov function (LF) $V : \mathbb{R}^n \rightarrow \mathbb{R}$ is used to analyze the asymptotic stability of the equilibrium. An LF for \eqref{eq: closed_loop_system} satisfies the following conditions:
%
\begin{subequations}
\begin{align}
    &V(\boldsymbol{0}_n) = 0,  \; V(\bfx) > 0, \; \forall \bfx \neq \boldsymbol{0}_n, \\  &\dot{V}(\bfx, \bfpi(\bfx)) = \nabla V(\bfx)^\top \bff_{\text{cl}}(\bfx)< 0, \; \forall \bfx \neq \boldsymbol{0}_n. \label{eq: LF_original_derivative_condition}
\end{align}
\label{eq: original_lf_conditions}
\end{subequations}
The existence of a radially unbounded ($V(\bfx) \to \infty$ as $\|\bfx\| \to \infty$) LF implies global asymptotic stability \cite{sastry2013nonlinear} of the dynamical system \eqref{eq: closed_loop_system}.

%

When the system model \eqref{eq: open_loop_control_system} is perfectly known, classical work~\cite{freeman_robust, SONTAG1989117} has shown how to synthesize stabilizing controllers when a control Lyapunov function is available. This has provided the theoretical basis for work aimed at learning Lyapunov-stable controllers \cite{jarvis_clf_sos, Chang2019NeuralLC, boffi2021learning}.
The availability of an LF-controller pair provides a complete solution, as the controller ensures system stabilization and the LF certifies its validity. However, real-world scenarios often present uncertainty in the dynamics model \eqref{eq: open_loop_control_system} that vary based on different physical parameters or operational conditions. Our goal is to find an LF-controller pair when uncertainty is present in the model. In particular, we do not assume any known error bounds or known distribution for the system uncertainty and instead investigate how to handle it based on finitely many realizations.

\section{Problem Formulation}\label{sec: problem_formulation}

We aim to synthesize a Lyapunov-stable controller for the continuous-time system with model uncertainty: 
\begin{equation}\label{eq: uncertain_system}
\dot{\bfx} = \bar{\bff}(\bfx, \bfu, \bfxi) := \bff(\bfx, \bfu) + \bfW(\bfx, \bfu) \bfxi,
\end{equation}
where $\bfx \in \calX \subset \bbR^n$ denotes the state vector, \(\bfu \in \calU \subset \mathbb{R}^m\) is the control input, and the functions \(\bff : \mathbb{R}^n \times \mathbb{R}^m \rightarrow \mathbb{R}^n\) and \(\bfW: \mathbb{R}^n \times \mathbb{R}^m \rightarrow \mathbb{R}^{n \times k} \) are locally Lipschitz. The set $\calX \subset \mathbb{R}^n$ represents the domain of interest, which includes the origin, and  $\calU \subset \mathbb{R}^m$ denotes the set of admissible control inputs. We assume $\calX$ and $\calU$ are compact.  

Each column of \(\bfW\) represents a specific direction of uncertainty (e.g., mass, friction). The vector \(\bfxi = [\xi_1, \xi_2, \ldots, \xi_k]^\top\) characterizes the uncertainty in the system model, and the distribution $\bbP^*$ of  $\bfxi$ is supported on a compact set $\Xi \subset \bbR^k$.
Additionally, we assume that $\bfW(\boldsymbol{0}_n, \boldsymbol{0}_m) = \boldsymbol{0}_{n \times k}$, which  ensures the origin is the desired equilibrium. 

\begin{figure}[htb]
  \centering
  \includegraphics[width=\linewidth]{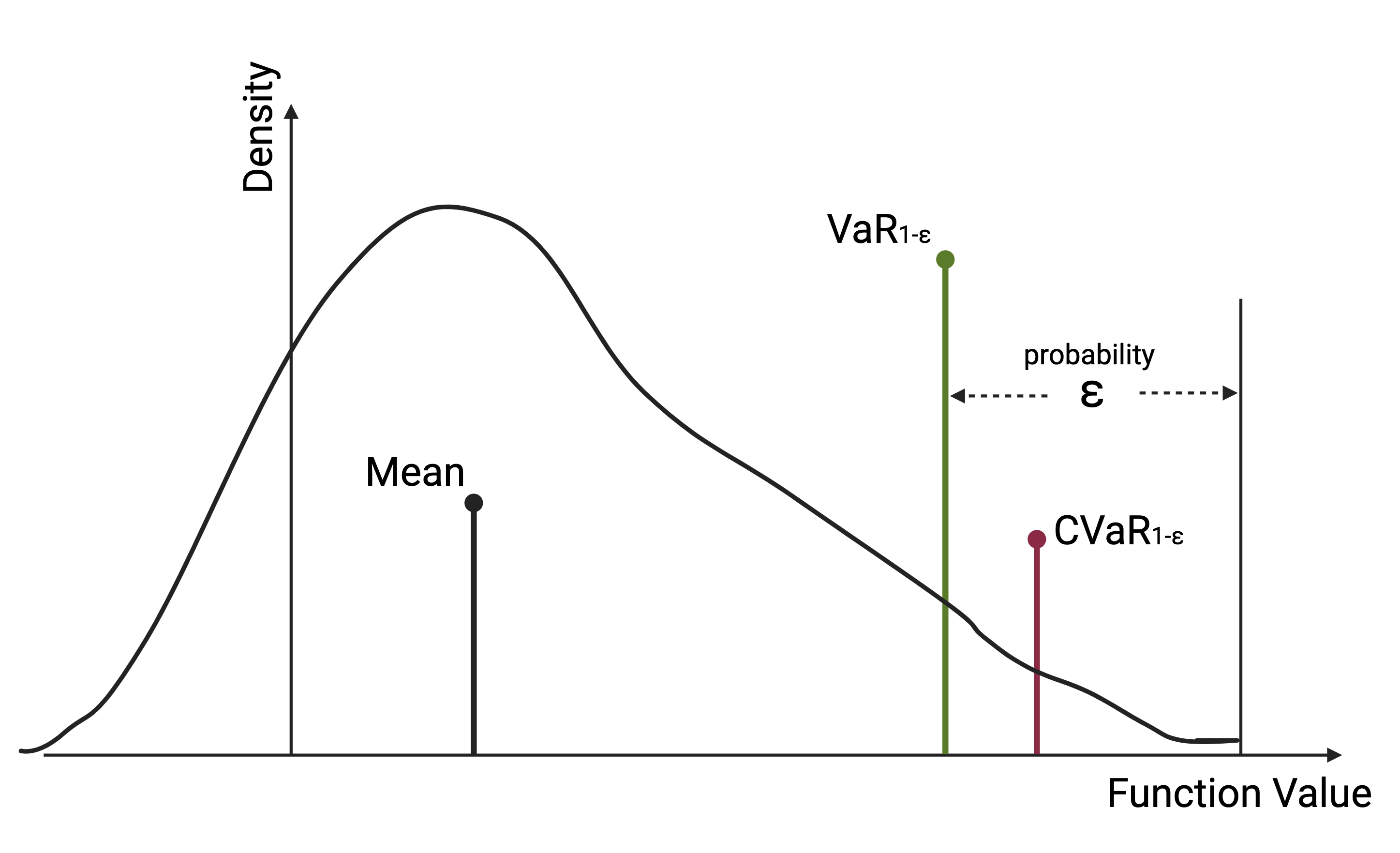}
  \caption{Illustration of VaR and CVaR within the distribution of a random variable. $\textrm{VaR}_{1-\epsilon}$ is the lower $1-\epsilon$ percentile of the random variable while $\textrm{CVaR}_{1-\epsilon}$ computes the expected value of the realizations above VaR.}
  \label{fig:var_and_cvar}
  \vspace*{-1ex}
\end{figure}

With the aim of stabilizing a system with model uncertainty in~\eqref{eq: uncertain_system}, we review chance-constraint formulations and their convex approximations. The Conditional Value-at-Risk (CVaR) is particularly useful in this context. For a random variable $Q$ with distribution $\bbP_q$, the Value-at-Risk (VaR) at confidence level $1 - \epsilon$ is defined as $\textrm{VaR}_{1-\epsilon}^{\mathbb{P}_q}(Q) := \inf_{t \in \bbR}\{t \; | \; \bbP_q(Q \leq t) \geq 1 - \epsilon\}$. Based on VaR, CVaR is defined~\cite{Rockafellar00optimizationof}  as $\textrm{CVaR}_{1-\epsilon}^{\mathbb{P}_q}(Q) = \bbE_{\mathbb{P}_q} [ Q \; | \; Q \geq \textrm{VaR}_{1-\epsilon}^{\mathbb{P}_q}(Q)] $.  Fig.~\ref{fig:var_and_cvar} shows an illustration of VaR and CVaR.


Our goal is to find a function $V: \mathbb{R}^n \mapsto \mathbb{R}$ and a controller $\bfpi: \bbR^n \mapsto \bbR^m$ that satisfy the Lyapunov conditions in~\eqref{eq: original_lf_conditions}. 
However, the uncertainty in the dynamical system \eqref{eq: uncertain_system}, which manifests itself in the term $\dot{V}(\bfx,\bfpi(\bfx))$, poses a challenge for ensuring that the condition on the derivative of the Lyapunov function~\eqref{eq: LF_original_derivative_condition} are satisfied. 
Therefore, we consider the chance-constrained condition,
%
\begin{equation}
\label{eq: ccp}
\bbP^*( \sup_{\bfx \in \calX} (\dot{V}(\bfx, \bfpi(\bfx), \bfxi) + \gamma \| \bfx \|) \leq 0) \geq 1-\epsilon,
\end{equation}
where $\gamma \in \bbR_{>0}$ and $\epsilon \in (0,1)$ is a user-specified risk tolerance. The chance constraint \eqref{eq: ccp} requires that, with probability at least $1-\epsilon$, the function $\dot{V}(\bfx, \bfpi(\bfx), \bfxi) + \gamma \|\bfx\|$ is less than or equal to zero for all $\bfx \in \calX$. This condition ensures that the Lyapunov derivative is negative definite with high probability.
%
%
%

The feasible set of the probabilistic constraint~\eqref{eq: ccp} is generally non-convex. To address this, Nemirosvski and Shapiro \cite{Nemirovski2006ConvexAO} proposed a convex CVaR approximation of the chance constraint in \eqref{eq: ccp} as follows, 
\begin{equation}
\label{eq: cvar_ccp}
\begin{aligned}
     \textrm{CVaR}_{1-\epsilon}^{\mathbb{P}^*}(\sup_{\bfx \in \calX} (\dot{V}(\bfx, \bfpi(\bfx), \bfxi) + \gamma \| \bfx \|)) \leq 0 .
\end{aligned}
\end{equation}
This inequality implies the satisfaction of the original chance constraint in~\eqref{eq: ccp}. As shown in~\cite{Rockafellar00optimizationof}, the inequality~\eqref{eq: cvar_ccp} can be written equivalently as:
\begin{align}
\label{eq: cvar_opti_def}   
    \inf_{t \in \mathbb{R}}[\epsilon^{-1}\mathbb{E}_{\mathbb{P}^*}[(\sup_{\bfx \in \calX} (\dot{V}(\bfx, \bfpi(\bfx), \bfxi) + \gamma \| \bfx \|)+t)_+]-t] \leq 0.
\end{align}
Using the formulations in \eqref{eq: ccp} and \eqref{eq: cvar_ccp} requires knowledge of the distribution $\bbP^*$.
However, in robotics and control applications typically $\bbP^*$ is unknown but a limited number of samples \(\{\bfxi_i\}_{i \in [N]}\) of the uncertain system parameters is available instead. For example, these samples can be obtained by collecting state-action sequences from a real system or a high-fidelity simulation and comparing the observed states with the predicted states based on a nominal model. The differences between the true and estimated states can be used as samples of the uncertain model parameters. Moreover, the distribution of $\bfxi$ may shift at system deployment time. This motivates us to consider the following problem.

\begin{problem}[\textbf{Distributionally robust Lyapunov function and controller learning}]\label{prob: DR_Controller}
Consider the system \eqref{eq: uncertain_system} with nominal dynamics $\bff$ and perturbation \(\bfW\). 
Let \(\{\bfxi_i\}_{i \in [N]}\) be finitely many samples of the uncertainty \(\bfxi\). Taking potential distribution shift into account, design a control policy \(\bfpi: \mathbb{R}^n \rightarrow \mathbb{R}^m \) along with a Lyapunov certificate \(V: \mathbb{R}^n \rightarrow \mathbb{R}\) that ensures that the equilibrium of the closed-loop system is globally asymptotically stable with high probability.
\end{problem}

\section{Convex Distributionally Robust Stability Formulation}
\label{sec: dr_clf_formulation}

In this section, we develop a distributionally robust formulation for control synthesis and analyze the stability of the resulting closed-loop systems. Our approach builds upon the concepts of distributionally robust optimization \cite{Esfahani2018DatadrivenDR, Xie2021OnDR}.


\subsection{Distributionally Robust Lyapunov-Stable Constraint}
\label{sec: prelim_drcc}


Let $\mathbb{P}_N :=  \frac{1}{N}\sum_{i=1}^N \delta_{\bfxi_i}$
denote the discrete empirical distribution obtained from the samples $\{\bfxi_i\}_{i \in [N]}$. We consider an ambiguity set of possible distributions for $\bfxi$, which are similar to the empirical distribution $\bbP_N$. Let $\calP_p(\Xi) \subseteq \calP(\Xi)$ be the set of Borel probability measures with finite $p$-th moment for $p \geq 1$. The $p$-Wasserstein distance between two probability measures $\mu$, $\nu$ in $\calP_p(\Xi)$ is defined as \cite{Xie2021OnDR}:
\begin{equation}
\label{eq: wasserstein_def}
    W_{p}(\mu,\nu) := \left(\inf_{\gamma \in \bbQ(\mu,\nu)} \left[ \int_{\Xi \times \Xi} \|x-y\|^p \text{d}\gamma(x,y) 
    \right] \right)^{\frac{1}{p}}, 
\end{equation}
where $\bbQ(\mu,\nu)$ is the set of measures on $\Xi \times \Xi$ with marginals $\mu$ and $\nu$ on the first and second factors. We define an ambiguity set 
      $\calM_{N}^{r} := \{\mu \in \calP_p(\Xi) \; | \; W_p(\mu,\mathbb{P}_{N} ) \leq r\}$
as a Wasserstein ball of distributions centered at $\mathbb{P}_N$ with radius~$r$. 

\begin{remark}[\textbf{Choice of Wasserstein ball radius}]\label{rem:choice-r}
{\rm
There is a relationship between the sample size $N$ and the Wasserstein radius $r$ for constructing the ambiguity set $\calM_N^r$. A distribution $\mathbb{P}$ is light-tailed if there exists an exponent $\rho$ such that $C := \mathbb{E}_{\mathbb{P}}[\exp{\norm{\boldsymbol{\xi}}^\rho}]=\int_{\Xi}\exp{\norm{\boldsymbol{\xi}}^\rho}\mathbb{P}(d\boldsymbol{\xi})<\infty$.
If the true distribution $\mathbb{P}^*$ of $\bfxi$ is light-tailed, the choice of radius $r=r_N(\bar{\epsilon})$ given in \cite[Theorem 3.5]{Esfahani2018DatadrivenDR},
    \begin{align}
    \label{eq: wasserstein_r_guarantee}
        r_N(\bar{\epsilon}) = \begin{cases}
            (\frac{\log(c_1\bar{\epsilon}^{-1})}{c_2 N})^{\frac{1}{\max\{k,2\}}} \quad &\text{if} \ N\geq \frac{\log(c_1\bar{\epsilon}^{-1})}{c_2}, \\
            (\frac{\log(c_1\bar{\epsilon}^{-1})}{c_2 N})^{\frac{1}{\rho}} \quad &\text{else},
        \end{cases}
    \end{align}
    where $c_1$ and  $c_2$ are
    positive constants that depend on $\rho, C$ and $k$ (cf.~\cite[Theorem 3.4]{Esfahani2018DatadrivenDR}), ensures that the ambiguity set $\calM_N^{r_{N}(\bar{\epsilon})}$ contains $\mathbb{P}^*$ with probability at least $1-\bar{\epsilon}$. 
    } \demo
\end{remark}

To account for the potential discrepancy between the empirical distribution $\mathbb{P}_N$ and the true distribution $\mathbb{P}^*$ of the system uncertainty $\bfxi$ at run time, we aim to find a pair $(V^*,\bfpi^*)$ that satisfies the following distributionally robust Lyapunov derivative constraint,
\begin{equation}
\begin{aligned}
\label{eq: dr_clf_control_constraint}
     \inf_{\bbP \in \calM_{N}^{r}}\mathbb{P}(\sup_{\bfx \in \calX} (\dot{V}^*(\bfx, \bfpi^*(\bfx), \bfxi) + \gamma \| \bfx \|) \leq 0) \geq 1 - \epsilon . 
\end{aligned}
\end{equation}
%
Compared with the chance-constrained formulation in~\eqref{eq: ccp}, the distributionally robust chance-constrained formulation~\eqref{eq: dr_clf_control_constraint} requires only a finite set of samples instead of the true distribution $\bbP^*$. Moreover, it offers robust constraint satisfaction guarantees against potential shifts of the uncertainty distribution within the constructed ambiguity set.

\subsection{Distributionally Robust Stability Characterization}

Next, we characterize the stability properties of the closed-loop system governed by a controller satisfying~\eqref{eq: dr_clf_control_constraint}.  If a pair $(V^*,\bfpi^*)$ satisfies \eqref{eq: dr_clf_control_constraint}, the following result ensures that the closed-loop system satisfies a chance constraint under the true distribution.

\begin{lemma}[\textbf{Chance-constraint satisfaction under the true distribution}]\label{lemma: dr_controller_guarantee}
Assume the distribution $\bbP^*$ of $\bfxi$ in \eqref{eq: uncertain_system} is light-tailed and the Wasserstein radius $r_N(\bar{\epsilon})$ is set according to~\eqref{eq: wasserstein_r_guarantee}. If the controller $\bfpi^*(\bfx)$ and Lyapunov function $V^*(\bfx)$ pair satisfies \eqref{eq: dr_clf_control_constraint} with $r=r_N(\bar{\epsilon})$, then,
\begin{equation}
\label{eq: dr_LF_stable_guarantee}
\bbP^*(\sup_{\bfx \in \calX}(\dot{V}^*(\bfx,\bfpi^*(\bfx), \bfxi) + \gamma \| \bfx \|) \leq 0) \geq (1-\epsilon)(1- \bar{\epsilon}).     
\end{equation}
\end{lemma}

\begin{proof}
Let $A := \{\bbP^* \in \calM_N^{r_N(\bar{\epsilon})}\}$ be the event that the true distribution $\bbP^*$ lies within the ambiguity set $\calM_N^{r_N(\bar{\epsilon})}$, and let $B := \{\sup_{\bfx \in \calX}(\dot{V}^*(\bfx, \bfpi^*(\bfx), \bfxi) + \gamma \| \bfx \|) \leq 0\}$ be the event that the Lyapunov derivative condition holds. From \cite[Theorem 3.4]{Esfahani2018DatadrivenDR}, we have $\bbP^*(A) \geq 1- \bar{\epsilon}$. From \eqref{eq: dr_clf_control_constraint}, we have that
\begin{equation}
\label{eq: dr_condition_x}
\inf_{\mathbb{P}\in \calM_{N}^{r_N(\bar{\epsilon})}}\bbP(B) \geq 1-\epsilon. \notag
\end{equation}
Now, consider the probability of the event $B$ under the true distribution $\bbP^*$:
\begin{align}
\bbP^*(B) &\geq \bbP^*(B \cap A) = \bbP^*(B | A)\bbP^*(A) \notag \\
&\geq \left(\inf_{\bbP \in \calM_{N}^{r_N(\bar{\epsilon})}}\bbP(B)\right)\bbP^*(A) \geq (1-\epsilon)(1-\bar{\epsilon}) \qedhere \notag
\end{align}
%
\end{proof}

According to Lemma~\ref{lemma: dr_controller_guarantee}, the closed-loop system satisfies the Lyapunov stability conditions pointwise in the state space with high probability. However, to analyze the global stability of the closed-loop system, it is necessary to extend this guarantee to the trajectories of the system over time. To this end, the following result links pointwise Lyapunov stability to asymptotic stability in probability. 

\begin{lemma}[\textbf{Global asymptotic stability in probability}]
\label{lemma: gloabl_stability_probability}
Let \(V^*: \mathbb{R}^n \rightarrow \mathbb{R}\) be a positive definite function with \(V^*(\boldsymbol{0}_n) = 0\), and let $V^*$ and the controller $\bfpi^*: \bbR^n \mapsto \bbR^m$ be a pair satisfying \eqref{eq: dr_LF_stable_guarantee}. Suppose $\bfxi \sim \bbP^*$, then, the origin $\boldsymbol{0}_n$  of the closed-loop system $\dot{\bfx} = \bar{\bff}(\bfx, \bfpi^*(\bfx), \bfxi)$ is globally asymptotically stable with probability at least $(1 - \epsilon)(1-\bar{\epsilon})$.
\end{lemma}

\begin{proof}
Define the sets 
\begin{align*}
\calA &:= \{\bfxi \mid \sup_{\bfx \in \calX} (\dot{V}^*(\bfx, \bfpi^*(\bfx), \bfxi) + \gamma \| \bfx \| ) \leq 0 \}, 
\\
\calB &:= \{\bfxi \mid \text{the system is asymptotically stable at } \bfx = \boldsymbol{0}_n \}.
\end{align*}
For $\bfxi \in \calA$, we have \(\dot{V}^*(\bfx, \bfpi^*(\bfx), \bfxi) \leq -\gamma \| \bfx \|\), implying \(V^*\) decreases along the system trajectories, and leading to convergence to the equilibrium state \(\bfx = \boldsymbol{0}\).
Therefore, \(\calA \subseteq \calB\), and we have \(\bbP^*(\calB) \geq \bbP^*(\calA) \geq (1 - \epsilon)(1- \bar{\epsilon})\), which concludes the result.
\end{proof}

Lemmas \ref{lemma: dr_controller_guarantee} and \ref{lemma: gloabl_stability_probability} establish global asymptotic stability in probability of the closed-loop system. Lemma~\ref{lemma: dr_controller_guarantee} demonstrates that if a pair $(V^*, \bfpi^*)$ satisfies the distributionally robust derivative constraint \eqref{eq: dr_clf_control_constraint}, then the system fulfills the Lyapunov stability condition with high probability under the true distribution $\bbP^*$. Subsequently, Lemma~\ref{lemma: gloabl_stability_probability} formalizes the global asymptotic stability of the system. Therefore, the satisfaction of the constraint in \eqref{eq: dr_clf_control_constraint} ensures global asymptotic stability in probability.

\begin{remark}[\textbf{Exponential stability under additional conditions}]
\label{rm: exponential_stability}
{\rm 
Under additional conditions, Lemma~\ref{lemma: gloabl_stability_probability} can ensure global exponential stability in probability. Consider the closed-loop system $\dot{\bfx} = \bff(\bfx, \bfpi^*(\bfx), \bfxi)$ with equilibrium at $\boldsymbol{0}_n$. Suppose there exists a positive constant $\alpha$ such that the pair $(V^*, \bfpi^*)$ satisfies
\begin{align}
\label{eq: dr_clf_exp_constraint}
    \inf_{\mathbb{P}\in \calM_{N}^{r}}\bbP(\sup_{\bfx \in \calX}(\dot{V}^*(\bfx, \bfpi^*(\bfx), \bfxi) &+ \gamma \| \bfx \| + \alpha V^*(\bfx)) < 0) \notag \\
    &\geq 1 - \epsilon,
\end{align}
and $V^*$ satisfies
\begin{equation}
    \alpha_1 \| \bfx\|^p \leq V^*(\bfx) \leq \alpha_2 \| \bfx\|^p,
\end{equation}
%
%
for some constants $\alpha_1, \alpha_2, p > 0$. Similar to the reasoning presented in Lemma~\ref{lemma: dr_controller_guarantee} and \ref{lemma: gloabl_stability_probability}, we conclude that the equilibrium of the closed-loop system is globally exponentially stable~\cite{sastry2013nonlinear} with probability at least $(1 - \epsilon)(1- \bar{\epsilon})$.
} \demo
\end{remark}

\begin{remark}[\textbf{Connections to other probabilistic stability notions}]\label{rm: connections_probability_stability}
{\rm In the literature, there exist other notions of stability in probability, especially for stochastic discrete-time systems, see e.g.,~\cite{kushner1967stochastic, Teel_stochastic_stability, culbertson2023input}. 
For the stochastic discrete-time system:
\begin{equation}
\label{eq: stochastic_system}
  \bfx_{i+1} = \hat{\bff}(\bfx_i, \bfu_i, \bfxi_i),
\end{equation}
a popular stability notion refers to the probability that the state remains in a bounded region \cite[Definition 6]{culbertson2023input}. Formally, the system \eqref{eq: stochastic_system} is bounded in probability for some $K \in \bbN$ if there exists $M > 0$ and $\epsilon \in (0,1)$ such that
\begin{equation}
\label{eq: bounded_in_probability}
\bbP^*\Big\{ \max_{k \leq K} \|\bfx_k \| \leq M  \Big\} \geq 1 - \epsilon.
\end{equation}
This notion is widely used for ensuring finite-time stability and safety~\cite{steinhardt2012finite, santoyo2021barrier} in stochastic systems. In typical stochastic systems, the parameter $\bfxi$ varies as time progresses, which is a more general and challenging setting. In contrast, as outlined in~\eqref{eq: uncertain_system}, our work focuses on scenarios where~$\bfxi$ remains fixed over time, allowing us to derive stronger convergence results. Lemma~\ref{lemma: gloabl_stability_probability} not only certifies stability in probability over finite-time horizons but also extends to global asymptotic stability guarantees in probability.
%
%
} \demo
\end{remark}

%
%

\subsection{Reformulation of Distributionally Robust Stability Constraint}
Synthesizing a pair $(V^*, \bfpi^*)$ that satisfies~\eqref{eq: dr_clf_control_constraint} is challenging because the constraint encompasses an infimum over a set of probability measures and a supremum over the state space~$\calX$. We leverage recent advances in distributionally robust optimization \cite{Esfahani2018DatadrivenDR, Hota2019DataDrivenCC} to identify a sufficient condition.

To simplify the notation, we define $h: \mathbb{R}^k \rightarrow \mathbb{R}$:
\begin{equation}
\label{eq: h_xi_def}
h(\bfxi) := \sup_{\bfx \in \calX} \left( \dot{V}(\bfx, \bfpi(\bfx), \bfxi) + \gamma \| \bfx \| \right).
\end{equation}

Note that by~\eqref{eq: uncertain_system}, we have
\begin{equation}
\label{eq: V_dot_uncertain}
\dot{V}(\bfx, \bfpi(\bfx), \bfxi) = \nabla V(\bfx)^\top(\bff(\bfx, \bfpi(\bfx)) + \bfW(\bfx, \bfpi(\bfx))\bfxi) .
\end{equation}
As it is the supremum of affine functions of $\bfxi$, $h(\bfxi)$ is convex in $\bfxi$. Furthermore, as $\calX$ is compact, $V$ is continuously differentiable, and $\bff$ and $\bfW$ are locally Lipschitz, the supremum in \eqref{eq: h_xi_def} exists and is finite for each~$\bfxi$.


We present a sufficient condition for the distributionally robust Lyapunov derivative constraint in \eqref{eq: dr_clf_control_constraint} that relies on samples of $\bfxi$. This condition provides a tractable reformulation of the constraint, making it more amenable to finding and learning the pair $(V^*, \bfpi^*)$.

\begin{proposition}[\textbf{Distributionally Robust Stability Condition}]
\label{proposition: relaxed_dr_clf_constraint}
Let $\{\bfxi_i\}_{i \in [N]}$ be a set of samples of the uncertainty parameter $\bfxi$, and let $h$ be the function defined in \eqref{eq: h_xi_def}. Assume that the samples are indexed such that $h(\bfxi_i) \geq h(\bfxi_k)$, for all $1 \leq i < k \leq N$. For $\epsilon \in (0,1)$, let $j \in [N]$ be the unique index that satisfies $\frac{j-1}{N} - \epsilon < 0$ and $\frac{j}{N} - \epsilon \geq 0$. Then, 
\begin{align}
\label{eq: dr_clf_constraint_relaxed}
&\frac{r}{\epsilon} \sup_{\bfx \in \calX} \| \bfW^\top(\bfx, \bfpi(\bfx)) \nabla V(\bfx) \| + \notag \\
&\qquad \frac{1}{N\epsilon}\sum_{i=1}^{j-1} (h(\bfxi_i) - h(\bfxi_j)) + h(\bfxi_j) \leq 0
\end{align}
is a sufficient condition for the distributionally robust stability constraint \eqref{eq: dr_clf_control_constraint} to hold for the ambiguity set corresponding to the 1-Wasserstein distance. Furthermore, if $\epsilon \leq \frac{1}{N}$, the condition \eqref{eq: dr_clf_constraint_relaxed} simplifies to:
\begin{align}
\label{eq: dr_clf_constraint_relaxed_simplified}
&\frac{r}{\epsilon} \sup_{\bfx \in \calX} \| \bfW^\top(\bfx, \bfpi(\bfx)) \nabla V(\bfx) \| + \max_{i} h(\bfxi_i) \leq 0.
\end{align}
\end{proposition}

\begin{proof}
Similar to the CVaR approximation~\eqref{eq: cvar_ccp} of the chance constraint~\eqref{eq: ccp}, the following is a sufficient condition for~\eqref{eq: dr_clf_control_constraint}: 
\begin{equation}
\begin{aligned}
\label{eq: drccp_cvar_approximate_initial}
    \sup_{\bbP \in \calM_{N}^{r}} \textrm{CVaR}_{1-\epsilon}^{\mathbb{P}}(h(\bfxi))) \leq 0 ,
\end{aligned}
\end{equation}
Using~\eqref{eq: cvar_opti_def}, we write~\eqref{eq: drccp_cvar_approximate_initial} equivalently as, 
\begin{equation}
\label{eq: prop_1_eq1}
    \sup_{\bbP \in \calM_{N}^{r}}\inf_{t \in \mathbb{R}}[\mathbb{E}_{\bbP}[(h(\bfxi) +t)_{+}] -t \epsilon] \leq 0 .
\end{equation}
Based on \cite[Lemma V.8]{Hota2019DataDrivenCC} and \cite[Theorem 6.3]{Esfahani2018DatadrivenDR}, with the 1-Wasserstein distance, the supremum over the ambiguity set can be written conservatively as the sample average over the empirical distribution $\bbP_N$ and a regularization term:
\begin{equation}
\label{eq: dr_constraint_rewrite}
r L_h + \inf_{t \in \mathbb{R}}\left[\mathbb{E}_{\bbP_N}[(h(\bfxi) + t)_{+}] - t \epsilon \right] \leq 0 ,
\end{equation}
where $L_h$ is the Lipschitz constant of $h(\bfxi)$ in $\bfxi$. Based on \cite[Proposition 1.32]{weaver2018lipschitz}, the Lipschitz constant of the supremum of a family of Lipschitz functions is bounded by the supremum of their Lipschitz constants. As shown in \eqref{eq: V_dot_uncertain}, $\dot{V}(\bfx, \bfpi(\bfx), \bfxi)$ is affine in $\bfxi$ for each $\bfx$ and its Lipschitz constant is given by $\| \bfW^\top(\bfx, \bfpi(\bfx)) \nabla V(\bfx) \|$. Therefore,
\begin{equation}
\label{eq: L_h_define}
L_h = \sup_{\bfx \in \calX} \| \bfW^\top(\bfx, \bfpi(\bfx)) \nabla V(\bfx) \|.
\end{equation}
Next, we write the second term in \eqref{eq: dr_constraint_rewrite} as:
\begin{equation}
\label{eq: prop_1_eq2}
    \inf_{t \in \mathbb{R}}\left[\mathbb{E}_{\bbP_N}[h(\bfxi) + t)_{+}] \!-\! t \epsilon \right] \!=\! \inf_{t \in \mathbb{R}}\left[\frac{1}{N}\sum_{i=1}^N (h(\bfxi_i) + t)_+\! - \! t \epsilon \right] .
\end{equation}
By assumption, $h(\bfxi_i) \geq h(\bfxi_k)$, for all $1 \leq i < k \leq N$. Observe that the function $\frac{1}{N}\sum_{i=1}^N (h(\bfxi_i) + t)+ - t \epsilon$ is piecewise-linear in $t$ with $N+1$ intervals and $N$ breakpoints, given by $\{-h(\bfxi_i)\}_{i \in [N]}$ and the slope for the $i$-th interval is $\frac{i-1}{N} - \epsilon$. Thus, the optimal solution is $t^* = -h(\bfxi_j)$, where $j \in [N]$ satisfies $\frac{j-1}{N} - \epsilon < 0$ and $\frac{j}{N} - \epsilon \geq 0$. Substituting the optimal solution $t^*$ into~\eqref{eq: prop_1_eq2}, we get:
\begin{equation}
\label{eq: prop_1_eq3}
\frac{1}{N}\sum_{i=1}^{j-1} (h(\bfxi_i) - h(\bfxi_j)) + \epsilon h(\bfxi_j) .
\end{equation}
%
%
Therefore, we know that:
\begin{align}
\label{eq: dr_clf_constraint_relaxed_proof}
&r \sup_{\bfx \in \calX} \| \bfW^\top(\bfx, \bfpi(\bfx)) \nabla V(\bfx) \| + \notag \\
&\qquad \frac{1}{N}\sum_{i=1}^{j-1} (h(\bfxi_i) - h(\bfxi_j)) + \epsilon h(\bfxi_j) \leq 0
\end{align}
is equivalent to~\eqref{eq: dr_constraint_rewrite}, and thus sufficient for~\eqref{eq: dr_clf_control_constraint}.

Furthermore, if $\epsilon \leq \frac{1}{N}$, only the first interval has negative slope and~\eqref{eq: prop_1_eq3} can be written equivalently as $ \epsilon h (\bfxi_1) = \max_{i} \epsilon h(\bfxi_i)$. Thus, given that $\epsilon \leq \frac{1}{N}$, we conclude that \eqref{eq: dr_clf_constraint_relaxed_simplified} is sufficient for \eqref{eq: dr_clf_control_constraint}.
\end{proof}

Based on Proposition~\ref{proposition: relaxed_dr_clf_constraint}, our goal is to find a controller $\bfpi^*(\bfx)$ and Lyapunov function $V^*(\bfx)$ pair that fulfills condition~\eqref{eq: dr_clf_constraint_relaxed}. Once such a pair is determined, leveraging Lemmas~\ref{lemma: dr_controller_guarantee} and~\ref{lemma: gloabl_stability_probability}, we can ascertain with high confidence that the system~\eqref{eq: uncertain_system}, governed by $\bfpi^*(\bfx)$, exhibits (exponential) asymptotic stability of the equilibrium.

\section{Distributionally Robust Lyapunov-Stable Controller Learning}
\label{sec: clf_nn_search}

In this section, we present a learning-based approach for finding a distributionally robust controller $\bfpi^*(\bfx)$ and Lyapunov function $V^*(\bfx)$ pair satisfying \eqref{eq: dr_clf_constraint_relaxed}. We make the following assumptions. 


\begin{assumption}[\textbf{Lipschitz Continuity and Boundedness}]\label{assump: system_dynamics_properties}
Assume the nominal system dynamics $\bff: \mathbb{R}^n \times \mathbb{R}^m \rightarrow \mathbb{R}^n$ and the perturbation function $\bfW: \mathbb{R}^n \times \mathbb{R}^m \rightarrow \mathbb{R}^{n \times k}$ are Lipschitz on $\calX \times \calU$ with constants $L_f$ and $L_W$, respectively, i.e., for all $\bfx, \bfx' \in \calX$ and $\bfu, \bfu' \in \calU$,
\begin{align*}
    \|\bff(\bfx, \bfu) - \bff(\bfx', \bfu')\| & \leq L_f (\|\bfx - \bfx'\| + \|\bfu - \bfu'\|),
    \\
    \|\bfW(\bfx, \bfu) - \bfW(\bfx', \bfu')\| & \leq L_W (\|\bfx - \bfx'\| + \|\bfu - \bfu'\|).
\end{align*}
Lipschitzness together with the compactness of the domain implies that both $\bff$ and $\bfW$ are uniformly bounded on $\calX \times \calU$. Specifically, there exist constants $B_f, B_W \in \mathbb{R}_{>0}$ such that, for any $\bfx \in \calX$ and $\bfu \in \calU$,
    \[
    \|\bff(\bfx, \bfu)\| \leq B_f \quad \text{and} \quad \|\bfW(\bfx, \bfu)\| \leq B_W .
    \]
\end{assumption}

The Lipschitz continuity assumption is particularly useful in our setting, as the neural networks are trained with discrete data, and we aim to certify stability for the entire state space. Lipschitzness ensures that the system dynamics do not change drastically for nearby states, which is crucial for the stability proofs presented later in this work.

\subsection{Model Architecture and Loss Function}
\label{sec: dr_model_loss}

To parameterize the desired pair $(V^*(\bfx), \bfpi^*(\bfx))$, we consider the following neural network representations. For the Lyapunov function, termed NN-LF, we set 
\begin{equation}
\label{eq: LF_NN_represent}
V_{\bftheta_1}(\bfx):= \| \phi_{\bftheta_1}(\bfx) -  \phi_{\bftheta_1}(\boldsymbol{0}_n) \|^2 + \hat{\alpha} \| \bfx \|^2,  
\end{equation} 
where $\phi_{\bftheta_1}: \bbR^n \mapsto \bbR$ is a fully-connected neural network with parameters $\bftheta_1$ and $\tanh$ activations, and $\hat{\alpha}$ is a user-chosen parameter \cite{dai_2021_lyapunov, gaby2021lyapunov_net}. By construction, this function is positive definite and $V_{\bftheta_1}(\boldsymbol{0}_n) = 0$. Furthermore, the use of $\tanh$ activations ensures continuous differentiability and bounded derivatives. Therefore, the gradient $\nabla V_{\bftheta_1}(\bfx)$ is Lipschitz with some constant $L_{\nabla V}$ and its norm $\|\nabla V_{\bftheta_1}(\bfx)\|$ is bounded within the compact set $\calX$ by some constant $B_V$. As shown in~\cite{fazlyab2019efficient}, these constants can be estimated explicitly, but we will only rely on their existence for stability analysis.

For the controller, we set
\begin{equation}
\label{eq: control_limit}
 \bfpi_{\bftheta_2}(\bfx) := \varphi_{\bftheta_2}(\bfx) - \varphi_{\bftheta_2}(\boldsymbol{0}_n),   
\end{equation}
where \(\varphi_{\bftheta_2}\) is a neural network with parameters \(\bftheta_2\) and \(\tanh\) activations. By construction, this formulation guarantees $\bfpi_{\bftheta_2}(\boldsymbol{0}_n) = \boldsymbol{0}_m$ and ensures that $\bfpi_{\bftheta_2}(\bfx)$ is locally Lipschitz in $\calX$, with constant $L_{\pi}$.

When employing a neural-network Lyapunov function, the Lyapunov derivative condition in~\eqref{eq: LF_original_derivative_condition} may not hold within a small neighborhood of the equilibrium due to numerical inaccuracies, see e.g.,~\cite{Chang2019NeuralLC, gaby2021lyapunov_net}. These inaccuracies stem from a neural network's inability to precisely approximate the derivative's behavior near the equilibrium, where its values are close to zero. This observation leads us to focus on establishing the validity of learned controller and Lyapunov function pairs outside of a neighborhood close to the origin. Therefore, our analysis concentrates on the region \(\mathcal{X}_{\delta} := \mathcal{X} \setminus \overline{B}(\boldsymbol{0}_n; \delta)\), which excludes a \(\delta\)-radius ball around the equilibrium. 

\begin{definition}[\textbf{\( \delta\)-accurate Lyapunov function over $\calX$} \cite{Chang2019NeuralLC}]
A Lyapunov function \(V(\bfx)\) for a dynamical system \(\dot{\bfx} = \bff(\bfx, \bfpi(\bfx))\) is \(\delta\)-accurate if it is positive definite, $V(\boldsymbol{0}_n) = 0$, and satisfies the Lyapunov derivative conditions~\eqref{eq: LF_original_derivative_condition} everywhere in $\calX_\delta$.  
\end{definition}

The existence of a \(\delta\)-accurate Lyapunov function implies the ultimate boundedness of the closed-loop system trajectories within a closed ball \( \overline{B}(\boldsymbol{0}_n;\delta)\). For a neural network representation, \(\delta > 0\) can be chosen arbitrarily small, potentially requiring a larger network size for \(V_{\bftheta_1}\) and $\bfpi_{\bftheta_2}$ and greater sampling density for the training set. For our distributionally robust formulation, we introduce a distributionally robust $\delta$-accurate Lyapunov function as follows.

\begin{definition}[\textbf{Distributionally robust \( \delta\)-accurate Lyapunov function with $\epsilon$ margin of error over $\calX$}]
A Lyapunov function \(V(\bfx)\) for a dynamical system \(\dot{\bfx} = \bar{\bff}(\bfx, \bfpi(\bfx), \bfxi)\) in~\eqref{eq: uncertain_system} is distributionally robust \(\delta\)-accurate if it is positive definite, $V(\boldsymbol{0}_n) = 0$, and satisfies:
\begin{equation*}
     \inf_{\bbP \in \calM_{N}^{r}}\mathbb{P}(\sup_{\bfx \in \calX_\delta} (\dot{V}(\bfx, \bfpi(\bfx), \bfxi) + \gamma \| \bfx \|) \leq 0) \geq 1 - \epsilon.
\end{equation*}
\end{definition}

Arguments analogous to the ones we employed in the proofs of Lemmas~\ref{lemma: dr_controller_guarantee} and~\ref{lemma: gloabl_stability_probability} show that the existence of a distributionally robust \(\delta\)-accurate Lyapunov function implies the stability of the closed-loop system to the closed ball \( \overline{B}(\boldsymbol{0}_n;\delta)\) with high probability.

To facilitate the training process of the pair $(V_{\bftheta_1}, \bfpi_{\bftheta_2})$, a training set $\calD_{\text{LF}} := \{ \bfx_i\}_{i \in [M]}$ is constructed by sampling $\calD_{\calX} := \{ \bfx_i\}_{i \in [M]}$ uniformly from $\calX_{\delta}$. We also assume the availability of the uncertainty training set $\calD_{\bfxi} := \{ \bfxi_i\}_{i \in [N]}$, collected by offline measurements of the system.

For a system without any uncertainty, we minimize the following empirical loss function that encourages the satisfaction of the Lyapunov derivative constraint~\eqref{eq: LF_original_derivative_condition}:
\begin{equation}
\label{eq: clf_loss}
    \ell_{\text{Nominal}}(\bftheta) = \frac{1}{M}\sum_{i=1}^M (\dot{V}_{\bftheta_1}(\bfx_i, \bfpi_{\bftheta_2}(\bfx_i)) + \gamma \|\bfx_i \|)_+,
\end{equation}
where $\bftheta = [\bftheta_1, \bftheta_2]$, and $\gamma \in \bbR_{> 0}$ is a user-chosen parameter.

To address the model uncertainty in~\eqref{eq: uncertain_system}, we aim to find a distributionally robust (DR)-NN LF $V_{\bftheta_1}$ and a controller $\bfpi_{\bftheta_2}$. We focus on the case where $\epsilon < \frac{1}{N}$ for two reasons. First, the DRO framework is particularly useful when the available data is limited, as it does not require a large number of samples to provide performance guarantees. In fact, the Wasserstein DRO approach is motivated by the fact that the true distribution is often only indirectly observable through a finite training dataset \cite{Esfahani2018DatadrivenDR}. When the number of samples $N$ is relatively small, the condition $\epsilon < \frac{1}{N}$ is also easy to satisfy, making it a practical choice for scenarios with limited data.
%
%
Second, setting $\epsilon < \frac{1}{N}$ ensures a higher probability of stability guarantees, as it requires the Lyapunov derivative condition to hold for all uncertainty samples in the training set $\calD_{\bfxi}$. Based on the stability condition in \eqref{eq: dr_clf_constraint_relaxed_simplified}, we consider the following loss function for learning a DR-NN LF and controller:
\begin{align}\label{eq: dr_lf_loss}
&\ell_{\text{DR}}(\bftheta) =  \bigg(\frac{r}{\epsilon} \max_{\bfx_i \in \calD_{\text{LF}}}\|\bfW^\top(\bfx_i, \bfpi_{\bftheta_2}(\bfx_i)) \nabla V_{\bftheta_1}(\bfx_i)\| + \notag
\\
& \max_{j}\big(\sum_{i = 1}^M (\dot{V}_{\bftheta_1}(\bfx_i, \bfpi_{\bftheta_2}(\bfx_i), \bfxi_j)) + \gamma \|\bfx_i \|)\big)\bigg)_+. 
\end{align}
%
%
%

In the following, for a sufficiently large $M$, we aim to show that the minimizer of the loss function~\eqref{eq: dr_lf_loss} yields a pair $(V_{\bftheta_1^*}, \bfpi_{\bftheta_2^*})$ satisfying the condition in~\eqref{eq: dr_clf_control_constraint}.

\subsection{Nominal Neural Lyapunov-Stable Control}
\label{sec: nominal_nn_lf}

First, we show based on results from~\cite{gaby2021lyapunov_net} that for a sufficiently large $M$, the nominal stabilizing controller learned from~\eqref{eq: clf_loss} stabilizes the nominal system. 
%
We formalize the necessary sampling density required to ensure stability using the learned LF.

\begin{lemma}(\cite[Lemma 4]{gaby2021lyapunov_net}).\label{lemma: covering_lemma}
For any \(\delta > 0\) and a chosen parameter $c > 0$, there exists a minimum
%
%
number of samples \(M(\delta, c) \in \mathbb{N}\) such that, for all $ M \geq M(\delta, c)$, a uniformly sampled training dataset \(\mathcal{D}_{\text{LF}} = \{\bfx_i\}_{i \in [M]} \subset \mathcal{X}_{\delta}\) ensures that the domain \(\mathcal{X}_{\delta}\) can be covered by the union of balls \(B(\bfx_i; c\|\bfx_i\|)\). Specifically, this guarantees that for any \(\bfx \in \mathcal{X}_{\delta}\), there exists \(\bfx_i \in \mathcal{D}_{\text{LF}}\) satisfying \(\|\bfx - \bfx_i\| \leq c\|\bfx_i\|\).
\end{lemma}
%
%

This result underscores the relation between \(\delta\), \(c\), and the sampling density \(M\), crucial for approximating the controller and Lyapunov function effectively. 

\begin{lemma}[\textbf{Neural Lyapunov-stable control}]\label{lemma: nominal_nn_lf}
Let $\mathcal{D}_{\text{LF}} = \{\bfx_i\}_{i \in [M]} \subset \mathcal{X}_{\delta}$ be a uniformly sampled training set, with $M \ge M(\delta, c)$ as defined in Lemma~\ref{lemma: covering_lemma}. Let $\boldsymbol{\theta}^* = (\boldsymbol{\theta}_1^*, \boldsymbol{\theta}_2^*)$ be the trained parameters such that $\ell_{\text{Nominal}}(\boldsymbol{\theta}^*) = 0$. Denote by $L_{\nabla V}$ and $L_{\pi}$ the Lipschitz constants of $\nabla V_{\boldsymbol{\theta}_1^*}$ and $\bfpi_{\boldsymbol{\theta}_2^*}$, respectively, and let $B_V$ be the bound on $\|\nabla V_{\boldsymbol{\theta}_1^*}(\bfx)\|$ for $\bfx \in \mathcal{X}$. Let $L_f$ and $B_f$ be as defined in Assumption~\ref{assump: system_dynamics_properties}. If $c > 0$ is sufficiently small to ensure $\gamma - (L_f(L_{\pi} + 1) B_V + L_{\nabla V} B_f)c > 0$, then the controller $\bfpi_{\boldsymbol{\theta}_2^*}$ stabilizes the nominal system $\bff$ to the closed ball $\overline{B}(\boldsymbol{0}_n;\delta)$ as certified by $V_{\boldsymbol{\theta}_1^*}$, which is a $\delta$-accurate Lyapunov function for the controlled system over~$\mathcal{X}$.
\end{lemma}

\begin{proof}
Given that \(\ell_{\text{Nominal}}(\boldsymbol{\theta}^*) = 0\), we have 
\begin{equation}\label{eq:samples-satisfy}
    \dot{V}_{\bftheta_1^*}(\bfx_i, \bfpi_{\bftheta_2^*}(\bfx_i))  + \gamma \|\bfx_i \| \leq 0 ,
\end{equation}
for all \( \bfx_i \in \mathcal{D}_{\text{LF}}\). 
Since \(\nabla V_{\bftheta_1^*}(\bfx)\) and $\bfpi_{\bftheta_2^*}$ are Lipschitz, based on the Assumption~\ref{assump: system_dynamics_properties}, we have for all $\bfx, \bfy \in \calX_{\delta}$,
\begin{align}\label{eq:lips-bounds}
\| \nabla V_{\bftheta_1^*}(\bfx) - \nabla V_{\bftheta_1^*}(\bfy) \| &  \leq L_{\nabla V} \| \bfx - \bfy \| , 
\\
\|\bff(\bfx, \bfpi_{\bftheta_2^*}(\bfx)) \! - \! \bff(\bfy, \bfpi_{\bftheta_2^*}(\bfy))\|  & \leq  
L_f(L_{\pi} \! + \! 1)\| \bfx \! - \! \bfy\|.  \notag 
\end{align}
We denote $L_q := L_f(L_{\pi} + 1)$ and write $\bff(\bfx, \bfpi_{\bftheta_2^*}(\bfx))$ as $\bff(\bfx)$ for brevity. Based on~\eqref{eq:lips-bounds}, we have for all $\bfx, \bfy \in \calX_{\delta}$,
\begin{align}
\label{eq: v_dot_lipschitz_nominal}
    &|\dot{V}_{\bftheta_1^*}(\bfx, \bfpi_{\bftheta_2^*}(\bfx)) - \dot{V}_{\bftheta_1^*}(\bfy, \bfpi_{\bftheta_2^*}(\bfy))| = \notag \\
    &|[\nabla V_{\bftheta_1^*}(\bfx)]^\top\bff(\bfx) - [\nabla V_{\bftheta_1^*}(\bfy)]^\top\bff(\bfy)| 
    \leq \notag \\
    &\|\nabla V_{\bftheta_1^*}(\bfx)\|\|\bff(\bfx) - \bff(\bfy)\| + \| \nabla V_{\bftheta_1^*}(\bfx) - \nabla V_{\bftheta_1^*}(\bfy)\| \|\bff(\bfy)\| \notag \\
    & \leq (L_q \| \nabla V_{\bftheta_1^*}(\bfx) \| +  L_{\nabla V} \| \bff(\bfy) \|) \| \bfx - \bfy\| \notag \\
    & \leq (L_q B_V +  L_{\nabla V} B_f)\| \bfx - \bfy\|.
\end{align}
This establishes that \(L_{\dot{V}} \), the Lipschitz constant for the Lyapunov derivative, can be bounded by \(L_q B_V +  L_{\nabla V} B_f\). By Lemma~\ref{lemma: covering_lemma}, for any \(\bfx \in \calX_{\delta}\), there exists a nearby sampled point \(\bfx_i\) such that \(\|\bfx - \bfx_i\| \leq c\|\bfx_i\|\).  Using~\eqref{eq:samples-satisfy} and~\eqref{eq: v_dot_lipschitz_nominal},
\begin{align}
\dot{V}_{\bftheta_1^*}(\bfx, \bfpi_{\bftheta_2^*}(\bfx)) &\leq \dot{V}_{\bftheta_1^*}(\bfx_i, \bfpi_{\bftheta_2^*}(\bfx_i)) + L_{\dot{V}}c\|\bfx_i\| \\
& \leq - \gamma \|\bfx_i \| + L_{\dot{V}}c\|\bfx_i\| \notag \\
& \leq - (\gamma - L_{\dot{V}}c) \delta. \notag
\end{align}
Since $\gamma - L_{\dot{V}}c > 0$, this ensures that \(\dot{V}_{\bftheta_1^*}(\bfx, \bfpi_{\bftheta_2^*}(\bfx))\) satisfies the Lyapunov derivative condition for all \(\bfx \in \mathcal{X}_{\delta}\), and thus it is a $\delta$-accurate Lyapunov function. Therefore, the learned controller provides \(\delta\)-accurate stabilization of the nominal system~$\bff$ as certified by the Lyapunov function.
\end{proof}

The stability assurances provided by Lemma~\ref{lemma: nominal_nn_lf} may not hold in the presence of the uncertainties in~\eqref{eq: uncertain_system}, which motivates our study of the system under the distributionally robust controller learned from~\eqref{eq: dr_lf_loss}.

\subsection{Neural Distributionally Robust Lyapunov-Stable Control}

For a sufficiently large sampling density, we establish probabilistic $\delta$-accurate stability guarantees for the uncertain system~\eqref{eq: uncertain_system} governed by the controller learned from~\eqref{eq: dr_lf_loss}.

\begin{proposition}[\textbf{Distributionally robust neural Lyapunov-stable control}]\label{proposition: dr_controller_LF}
Let $\mathcal{D}_{\text{LF}} = \{\bfx_i\}_{i \in [M]} \subset \mathcal{X}_{\delta}$ be a uniformly sampled training set, with $M \ge M(\delta, c)$ as defined in Lemma~\ref{lemma: covering_lemma}. Let $\boldsymbol{\theta}^* = (\boldsymbol{\theta}_1^*, \boldsymbol{\theta}_2^*)$ be the trained parameters such that $\ell_{\text{DR}}(\boldsymbol{\theta}^*) = 0$. Let $L_{\nabla V}$ and $L_{\pi}$ denote the Lipschitz constants of $\nabla V_{\boldsymbol{\theta}_1^*}$ and $\bfpi_{\boldsymbol{\theta}_2^*}$, respectively, and let $B_V$ be the bound on $\|\nabla V_{\boldsymbol{\theta}_1^*}(\bfx)\|$ for $\bfx \in \mathcal{X}$ and $\epsilon \leq \frac{1}{N}$. Let $c > 0$ be sufficiently small such that $\gamma - ((L_W(L_\pi + 1))B_V + L_{\nabla V}B_W)B_\xi \frac{r}{\epsilon}c - (L_f(L_{\pi} + 1) B_V + L_{\nabla V}B_f)c > 0$, where $L_W$, $L_f$, $B_W$, and $B_f$ are as defined in Assumption~\ref{assump: system_dynamics_properties}, and $B_\xi \in \bbR_{>0}$ is such that $\|\bfxi\| \leq B_\xi$ for all $\bfxi \in \Xi$. Then, the controller $\bfpi_{\bftheta_2^*}$ stabilizes the uncertain system~\eqref{eq: uncertain_system} to the closed ball $\overline{B}(\boldsymbol{0}_n;\delta)$ with high probability, as certified by $V_{\boldsymbol{\theta}_1^*}$, which is a distributionally robust $\delta$-accurate Lyapunov function for the controlled system over $\calX$.
\end{proposition}

\begin{proof}
Given that \(\ell_{\text{DR}}(\boldsymbol{\theta}^*) = 0\), we have for all \( \bfx_i \in \mathcal{D}_{\text{LF}}\), 
\begin{multline}
    \frac{r}{\epsilon} \max_{\bfx_i \in \calD_{\text{LF}}}\|\bfW^\top(\bfx_i, \bfpi_{\bftheta_2^*}(\bfx_i)) \nabla V_{\bftheta_1^*}(\bfx_i)\| + 
    \\
    \max_{j}(\dot{V}_{\bftheta_1^*}(\bfx_i, \bfpi_{\bftheta_2^*}(\bfx_i), \bfxi_j)) \notag + \gamma \|\bfx_i \| \leq 0.    
\end{multline}
Based on Assumption~\ref{assump: system_dynamics_properties} and the Lipschitz constant $L_{\pi}$ of the controller $\bfpi_{\bftheta_2^*}$,  we have for all $\bfx, \bfy \in \calX_{\delta}$:
\begin{equation}
\label{eq: W_lipschiz_proposition}
    \|\bfW(\bfx, \bfpi_{\bftheta_2^*}(\bfx)) - \bfW(\bfy, \bfpi_{\bftheta_2^*}(\bfx))\| \leq L_W(L_\pi + 1) \|\bfx - \bfy\|.
\end{equation}
For brevity, we denote $L_p = L_W(L_\pi + 1)$ and write $\bfW(\bfx, \bfpi_{\bftheta_2^*}(\bfx))$ and $V_{\bftheta_1^*}(\bfx)$ as $\bfW(\bfx)$ and $V(\bfx)$, respectively. 
Using~\eqref{eq:lips-bounds} and~\eqref{eq: W_lipschiz_proposition}, we have for all $ \bfx, \bfy \in \calX_{\delta}$: 
\begin{align}
\label{eq: w_dot_lipschitz}
&\|\bfW^\top(\bfx) \nabla V(\bfx) - \bfW^\top(\bfy) \nabla V(\bfy)\|  \leq \notag \\
&\|(\bfW^\top(\bfx) \!-\! \bfW^\top(\bfy)) \nabla V(\bfx)\| \!+\! \| \bfW^\top(\bfy) (\nabla V(\bfx) \!-\! \nabla V(\bfy)\| \notag \\
&\leq  (L_p \|\nabla V(\bfx) \| +L_{\nabla V}\|\bfW^\top(\bfy)\|)\|\bfx - \bfy\| \notag \\
&\leq  (L_p B_V  +L_{\nabla V} B_W)\|\bfx - \bfy\|.
\end{align}
Thus, the term $\bfW^\top(\bfx, \bfpi_{\bftheta_2^*}(\bfx)) \nabla V_{\bftheta_1^*}(\bfx)$ is Lipschitz with constant $L_{\dot{W}}$, bounded by $L_p B_V  +L_{\nabla V} B_W$.

Moreover, as $\bfxi \in \Xi$ and $\Xi$ is compact, we know that $\| \bfxi\| \leq B_{\xi}$ for some constant $B_{\xi} \in \bbR_{>0}$. Using~\eqref{eq: v_dot_lipschitz_nominal} and~\eqref{eq: W_lipschiz_proposition}, for the term \(\dot{V}_{\bftheta_1^*}(\bfx, \bfpi_{\bftheta_2^*}(\bfx), \bfxi)\), we have that, for each $\bfxi$, and for all $\bfx, \bfy \in \calX_{\delta}$,
\begin{align}
\label{eq: v_dot_lipschitz_uncertain}
&|\dot{V}_{\bftheta_1^*}(\bfx, \bfpi_{\bftheta_2^*}(\bfx), \bfxi) - \dot{V}_{\bftheta_1^*}(\bfy, \bfpi_{\bftheta_2^*}(\bfy), \bfxi)| \notag \\
& \; = |[\nabla V_{\bftheta_1^*}(\bfx)]^\top \bff(\bfx) - [\nabla V_{\bftheta_1^*}(\bfy)]^\top \bff(\bfy) \notag
\\
& \; \quad + ([\nabla V_{\bftheta_1^*}(\bfx)]^\top \bfW(\bfx) - [\nabla V_{\bftheta_1^*}(\bfy)]^\top \bfW(\bfy))\bfxi| \notag \\
& \; \leq L_{\dot{V}}\|\bfx - \bfy\| + L_{\dot{W}}\|\bfxi\| \|\bfx - \bfy\| \notag \\
& \; \leq (L_{\dot{V}} + L_{\dot{W}}B_{\xi}) \|\bfx - \bfy\|.
\end{align}
We denote $L_{\text{max}} := L_{\dot{V}} + L_{\dot{W}}B_{\xi}$. 
We define $G(\bfx) = \max_j \dot{V}_{\bftheta_1^*}(\bfx, \bfpi_{\bftheta_2^*}(\bfx), \bfxi_j)$. From~\eqref{eq: v_dot_lipschitz_uncertain}, we deduce 
%
%
$|G(\bfx) - G(\bfy)| \leq L_{\text{max}} \|\bfx - \bfy\|$, and thus $G$ is also Lipschitz with constant $L_{\text{max}}$.

Lastly, following the covering argument in Lemma~\ref{lemma: covering_lemma}, we have that for any $\bfx \in \calX_{\delta}$, there exists an \(\bfx_i \in \calD_{\text{LF}}\) such that \(\|\bfx - \bfx_i\| \leq c\|\bfx_i\|\). Therefore, $\forall \bfx \in \calX_{\delta}$,
\begin{align}
&\frac{r}{\epsilon} \sup_{\bfx \in \calX_{\delta}}\|\bfW^\top(\bfx) \nabla V(\bfx)\| + 
\max_{j}(\dot{V}_{\bftheta_1^*}(\bfx, \bfpi_{\bftheta_2^*}(\bfx), \bfxi_j)) \leq \notag \\
& \frac{r}{\epsilon} \max_{\bfx_i \in \calD_{\text{LF}}}\|\bfW^\top(\bfx_i) \nabla V(\bfx_i)\|+\max_{j}(\dot{V}_{\bftheta_1^*}(\bfx_i, \bfpi_{\bftheta_2^*}(\bfx_i), \bfxi_j)) + \notag \\
& \big(\frac{r}{\epsilon}L_{\dot{W}}c + L_{\text{max}}c\big)\|\bfx_i\| \leq - \big(\gamma - \frac{r}{\epsilon}L_{\dot{W}}c - L_{\text{max}}c\big) \delta.
\end{align}
Note that $\gamma - \frac{r}{\epsilon}L_{\dot{W}}c - L_{\text{max}}c > 0$.
%
%
Thus, we have that  $\frac{r}{\epsilon} \sup_{\bfx \in \calX_{\delta}}\| \bfW^\top(\bfx) \nabla V(\bfx) \| + \max_{j}(\dot{V} (\bfx, \bfpi(\bfx), \bfxi_j)) < 0 $ holds for all $\bfx \in \calX_{\delta}$. 

Therefore, reasoning as in Proposition~\ref{proposition: relaxed_dr_clf_constraint},  we conclude that, for any $\bfxi \sim \bbP^*$,~\eqref{eq: dr_clf_control_constraint} is satisfied for all $\bfx \in \calX_{\delta}$. This implies that $V_{\bftheta_1^*}$ is a distributionally robust $\delta$-accurate Lyapunov function. Therefore, the learned controller provides distributionally robust $\delta$-accurate stabilization of the uncertain system, as certified by the Lyapunov function.
\end{proof}


The universal approximation property of neural networks~\cite{hornik1989multilayer, leshno1993multilayer} suggests that, given sufficient network width and depth, the optimal Lyapunov function and controller can be approximated to arbitrary accuracy, making it possible to achieve zero training loss in the nominal~\eqref{eq: clf_loss} and distributionally robust~\eqref{eq: dr_lf_loss} settings.

The learning process for the DR-NN-LF and controller pair, as defined in~\eqref{eq: dr_lf_loss}, can be made more efficient by initializing the neural network weights with those obtained from the nominal training~\eqref{eq: clf_loss}. This initialization strategy significantly reduces the required training time and computational resources for the DR formulation.

\begin{remark}[\textbf{Neural verification of stability guarantees}]
{\rm
While the universal approximation property of neural networks suggests that the optimal Lyapunov function and controller can be approximated to arbitrary accuracy, achieving zero training loss may not always be possible in practice. In such cases, the stability guarantees provided by Lemma VI.4 and Proposition VI.5 may not hold exactly. However, similar to \cite{yang2024lyapunovstable}, by employing neural network verification techniques, such as alpha-beta-CROWN \cite{wang2021beta} and Marabou \cite{wu2024marabou}, one can formally verify whether the learned Lyapunov function and controller pair satisfy the required derivative conditions over a compact state set. } \demo
\end{remark}


\begin{remark}[\textbf{Pointwise distributionally robust formulation}]
{\rm
The formulation in Proposition \ref{proposition: relaxed_dr_clf_constraint} and the corresponding loss \eqref{eq: dr_lf_loss} may be conservative and challenging to train. The conservativeness arises from the term $\frac{r}{\epsilon} \max_{\bfx_i \in \calD_{\text{LF}}}\|\bfW^\top(\bfx_i, \bfpi_{\bftheta_2}(\bfx_i)) \nabla V_{\bftheta_1}(\bfx_i)\|$, which is a positive constant at any $\bfx_i$ in the training set $\calD_{\text{LF}}$. This term requires the Lyapunov derivative condition to be the negative of that constant for all training data, which can be a stringent requirement, especially for states near the origin.
%
%

An alternative approach is to consider a pointwise distributionally robust formulation, which enforces the chance constraint at each state $\bfx \in \calX$ individually:
\begin{equation}
\label{eq: point_ccp}
\bbP^*(\dot{V}(\bfx, \bfpi(\bfx), \bfxi) + \gamma \| \bfx \|  \leq 0) \geq 1-\epsilon, \; \; \forall \bfx \in \calX .
\end{equation}
Based on this pointwise formulation, we can define a pointwise loss function for training the neural network Lyapunov function and controller:
\begin{align}
\label{eq: point_dr_lf_loss}
\ell_{\text{DR}}(\bftheta) &= \frac{1}{M}\sum_{i=1}^M (\frac{r}{\epsilon} \|\bfW^\top(\bfx_i, \bfpi_{\bftheta_2}(\bfx_i)) \nabla V_{\bftheta_1}(\bfx_i)\| + \notag \\
& \quad \max_{j}(\dot{V}_{\bftheta_1}(\bfx_i, \bfpi_{\bftheta_2}(\bfx_i), \bfxi_j)) + \gamma \|\bfx_i \|)_+.
\end{align}
This pointwise loss function is more flexible and easier to train compared to the uniform loss~\eqref{eq: dr_lf_loss} because it allows for more variability in the Lyapunov derivative values across different states. By enforcing the distributionally robust chance constraint individually for each state, the pointwise formulation allows the learned Lyapunov function and controller to adapt to the specific characteristics of each state, potentially leading to faster training convergence.

%
%
This pointwise formulation would ensure that the Lyapunov function decreases with high probability at each individual state. However, providing a global stability guarantee based on this requires notions of stochastic stability \cite{culbertson2023input, Teel_stochastic_stability}. We leave the analysis of the stability properties of the closed-loop system under the controller learned using the pointwise loss function \eqref{eq: point_dr_lf_loss} for future work.
} \demo
\end{remark}




\section{Evaluation}
\label{sec: evaluation}



This section evaluates our approach for synthesizing distributionally robust Lyapunov-stable controllers. We focus on two classic control problems from Gymnasium \cite{towers_gymnasium_2023}: the Inverted Pendulum and the Continuous-time Mountain Car. For each system, we consider multiple instances with varying characteristics, each defined by its own physical parameters such as mass, length, and friction. These instances are used for training our distributionally robust Lyapunov-stable controller~\eqref{eq: dr_lf_loss}.

We use PyTorch~\cite{paszke2019pytorch} to train our neural network Lyapunov function and controller pair, both for the baseline case~\eqref{eq: clf_loss} and for the distributionally robust (DR) formulation~\eqref{eq: dr_lf_loss}. The models are trained using the Adam optimizer~\cite{Adam} with a learning rate of 0.002. The PyTorch clamp function is used to enforce input bounds for the control policy network, which is implemented with a smooth approximation and enables gradient back-propagation. The training was done on hardware with an Nvidia GeForce RTX 2070 Super GPU and an Intel i7 9700K CPU. 

During testing, we assume the physical parameters of each system may be drawn from distributions different from those of the training instances. This setup aims to assess the performance of the learned controller under realistic scenarios, where distributional shifts in the system parameters are common. By doing so, we effectively evaluate the robustness and adaptability of our controller in the presence of distributional uncertainty in the system models. 

\subsection{Inverted Pendulum}

\begin{figure}[t]
  \centering
  \subcaptionbox{Baseline Trajectories\label{fig:2a}}{\includegraphics[width=0.49\linewidth]{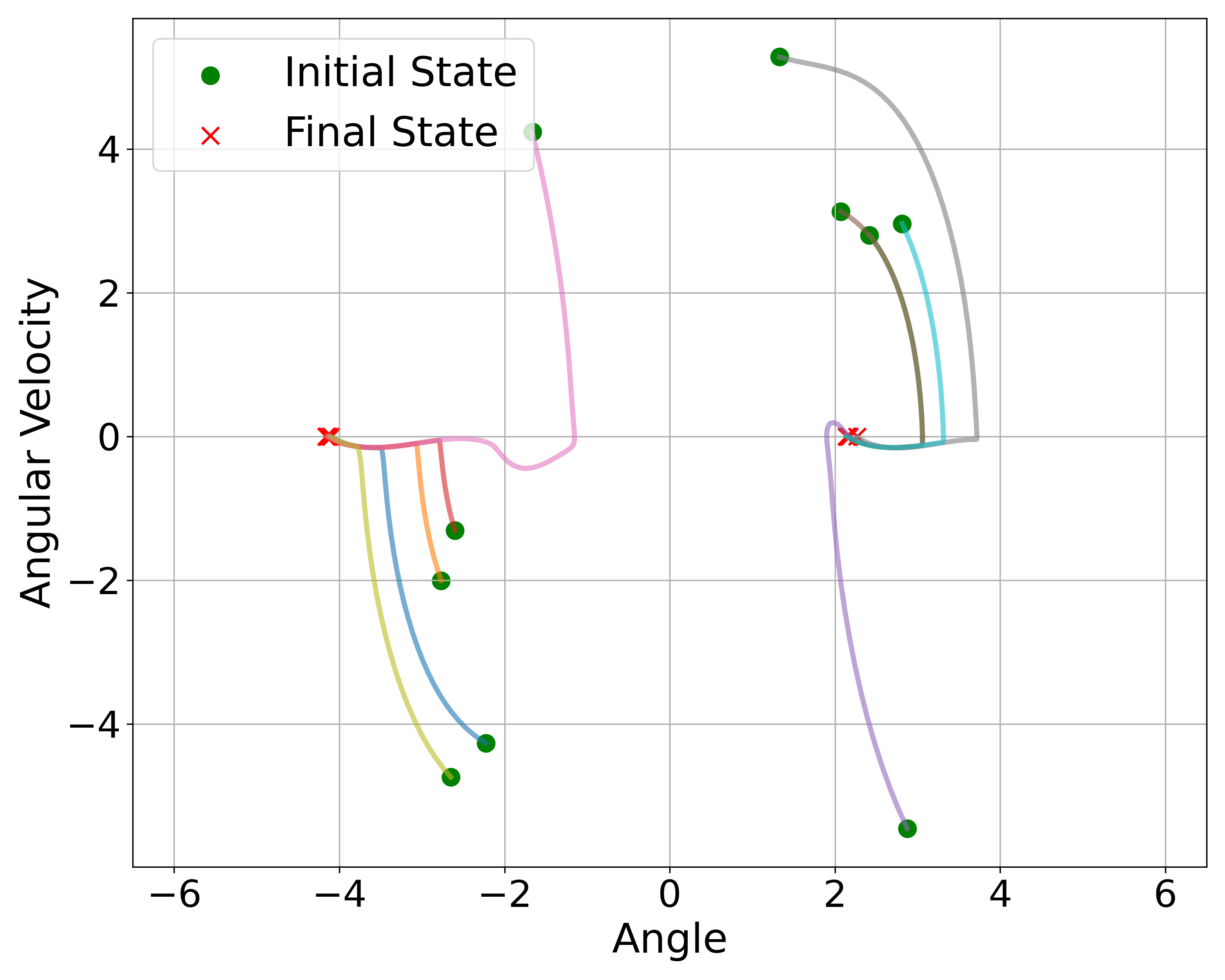}}%
  \hfill%
  \subcaptionbox{DR Trajectories\label{fig:2b}}{\includegraphics[width=0.49\linewidth]{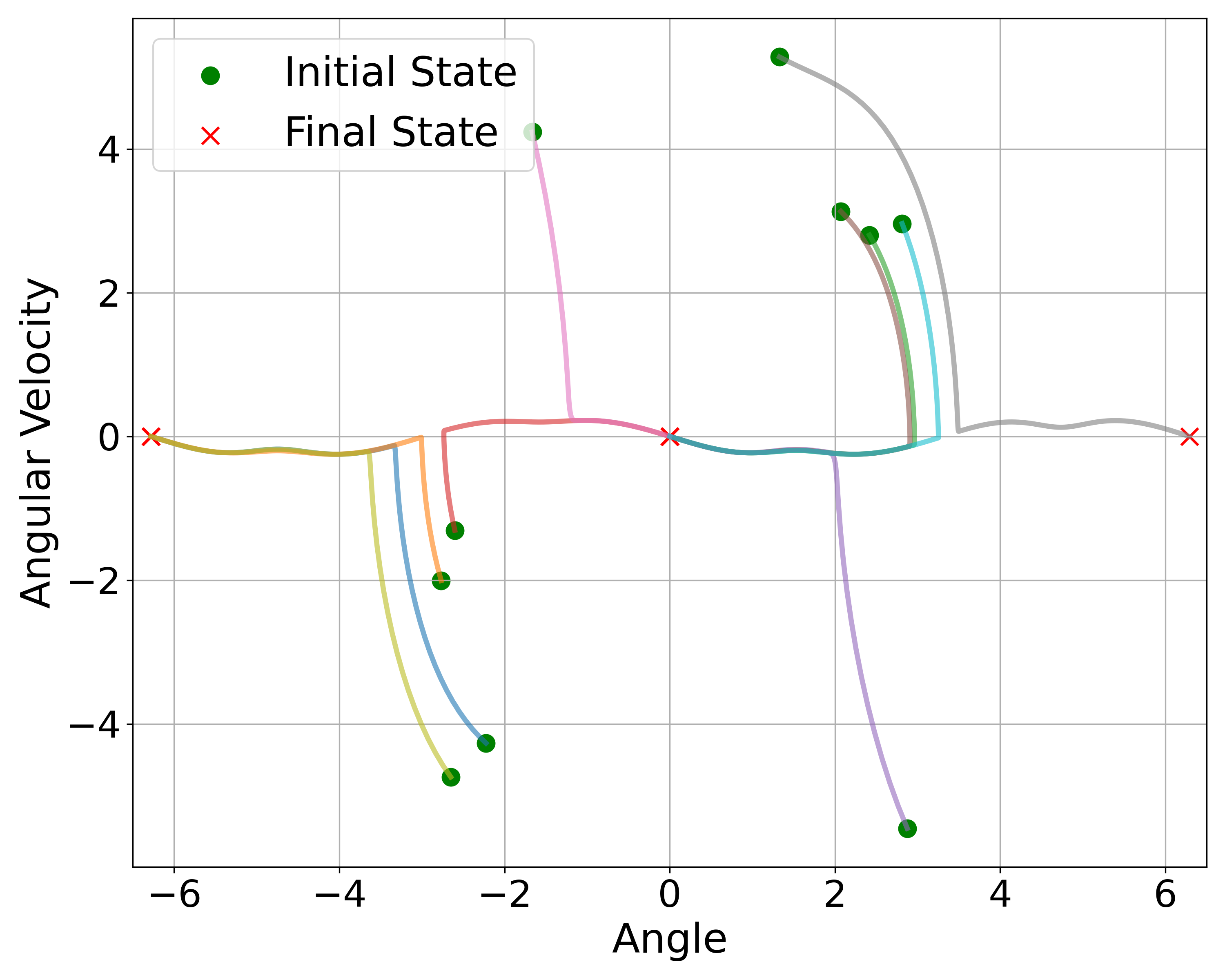}}\\
  \caption{Comparison of trajectories for the inverted pendulum system with test case parameters: mass $m=1.1$, length $l=1.0$, and damping $b=0.18$. The $10$ random sampled initial states are marked as green dots, while the final states are marked as red crosses. The states $(\theta, \dot{\theta}) = (2k\pi, 0)$ for $k \in \mathbb{N}$ are stable equilibrium states, while the points $((2k-1)\pi, 0)$ are unstable equilibrium points, corresponding to the upside-down position of the pendulum. In (a), the baseline controller, trained using the average mass and damping from offline observations, fails to stabilize the pendulum to upright. In (b), the distributionally robust (DR) controller successfully stabilizes the pendulum to an upright position for every initial state, demonstrating improved robustness to distributional shifts in the system parameters.}
  \label{fig: inverted_pendulum_traj}
\end{figure}

\begin{figure*}[ht]
  \centering
  \subcaptionbox{Lyapunov Function Values\label{fig:3a}}{\includegraphics[width=0.245\textwidth]{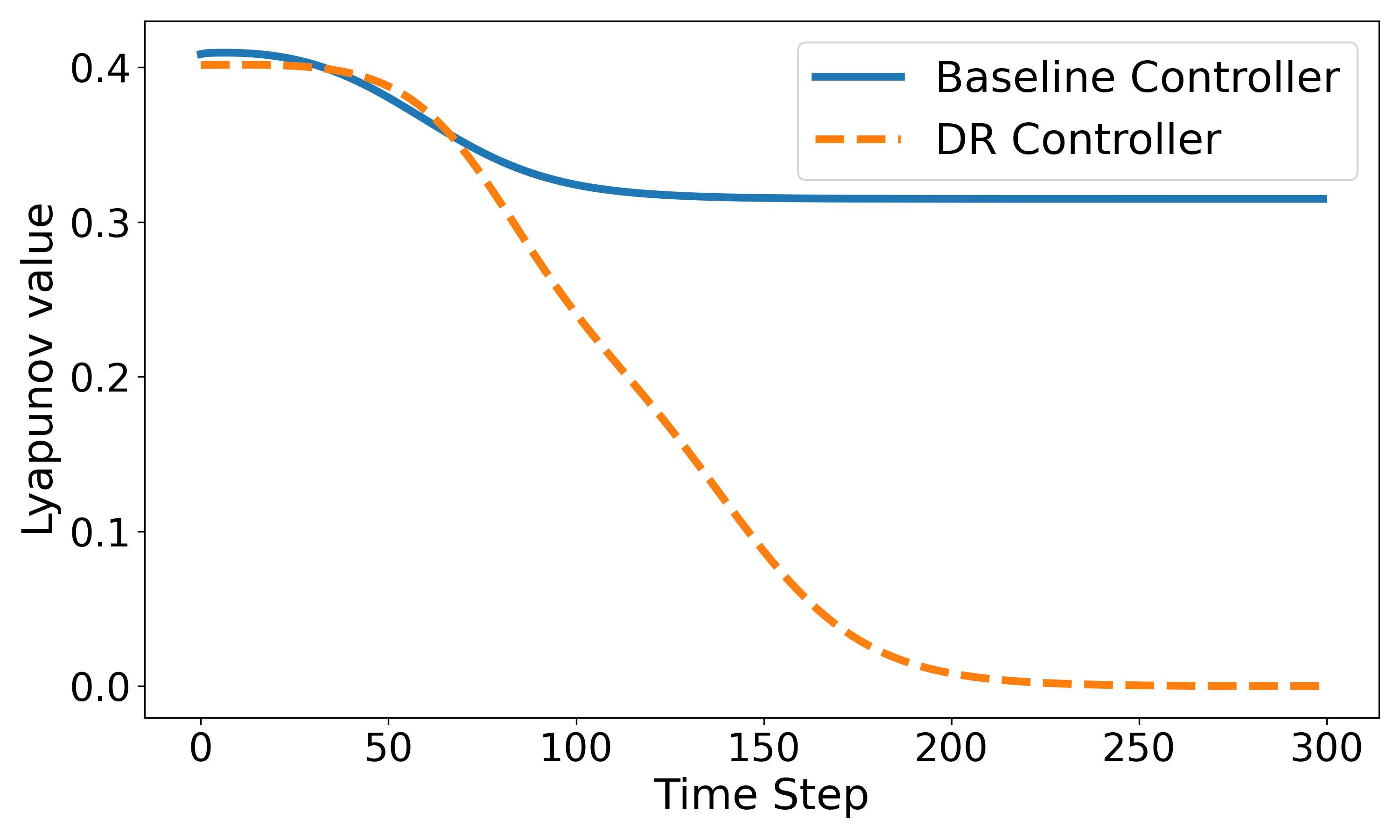}}%
  \hfill%
  \subcaptionbox{Control Inputs\label{fig:3b}}{\includegraphics[width=0.245\textwidth]{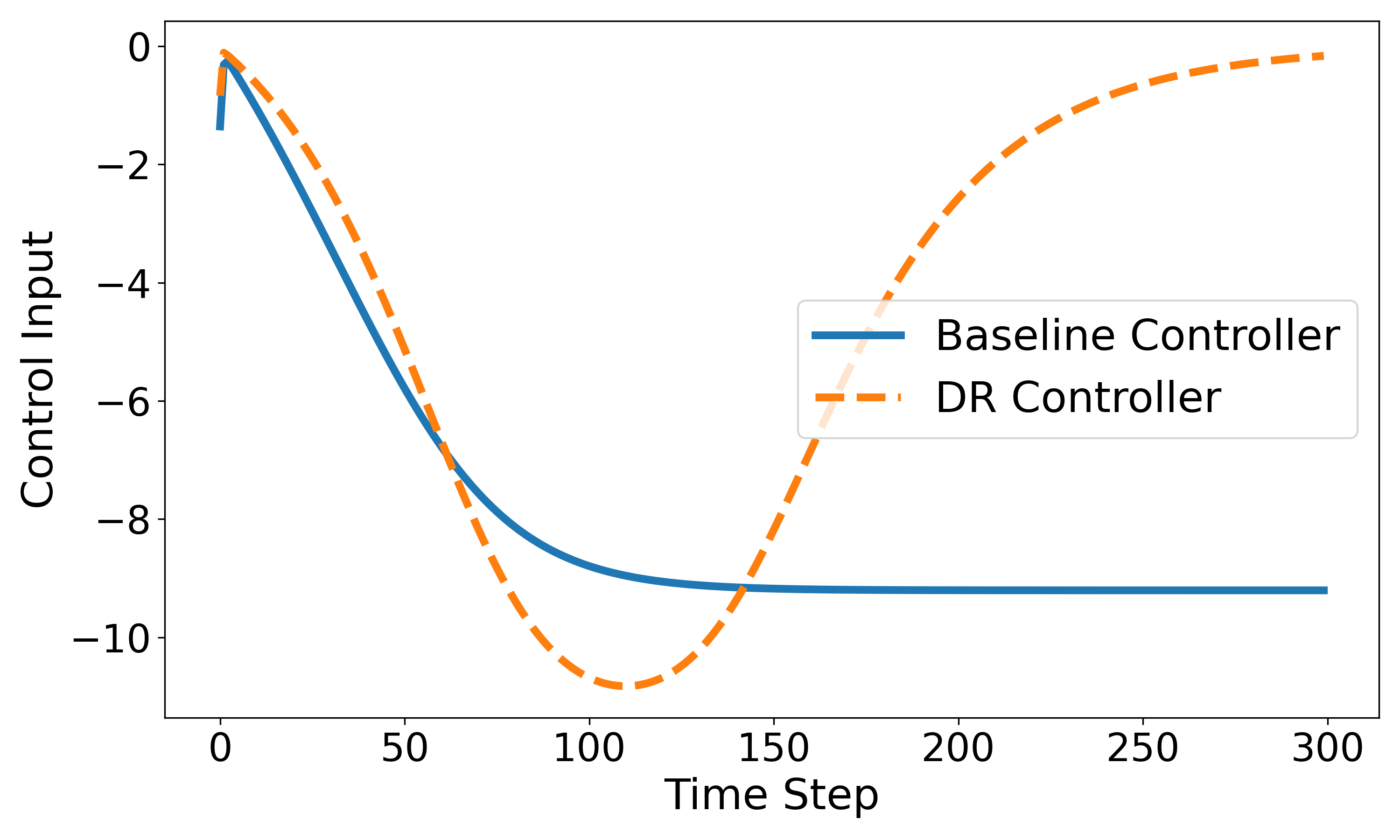}}
  \subcaptionbox{RL Value Functions\label{fig:3c}}{\includegraphics[width=0.245\textwidth]{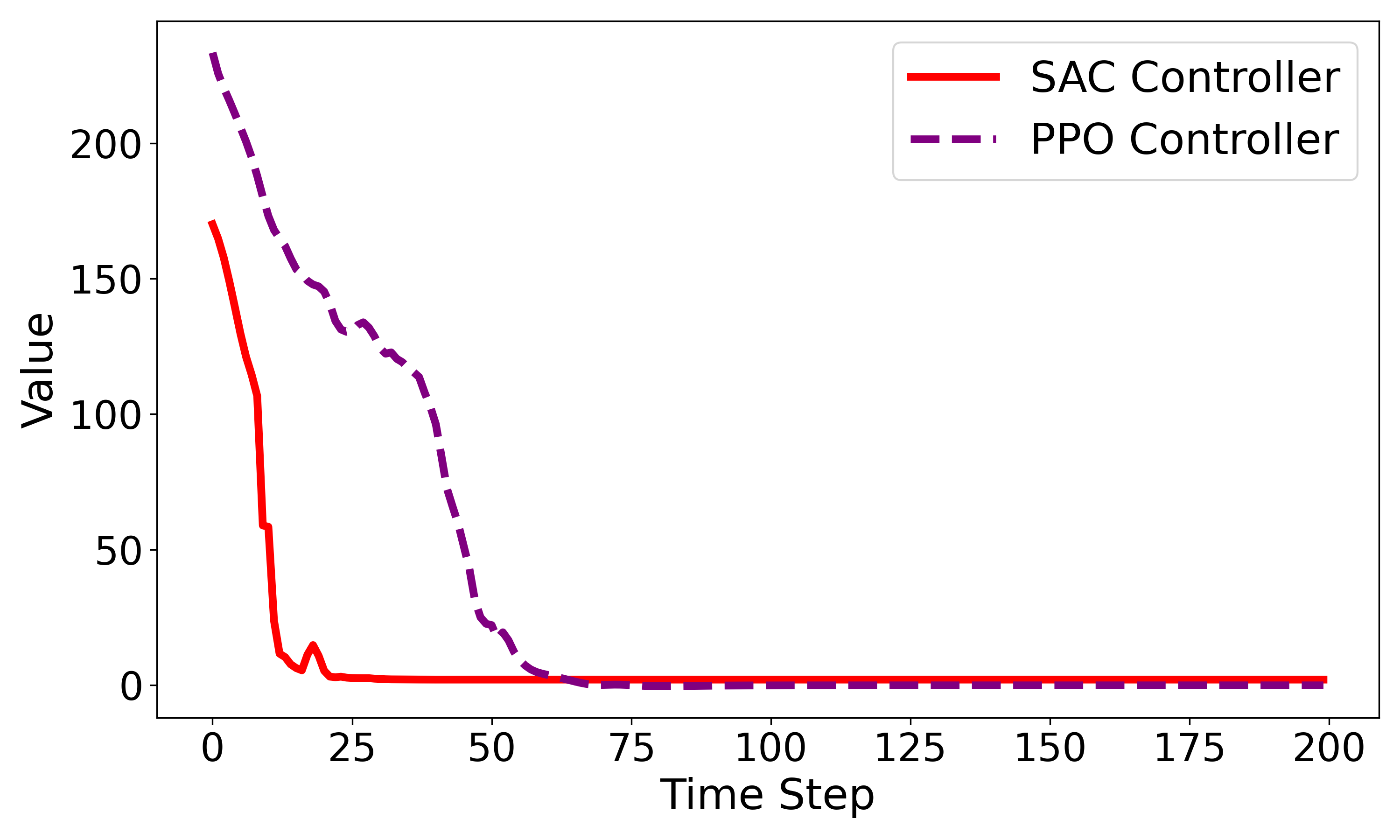}}%
  \hfill%
  \subcaptionbox{RL Control Inputs\label{fig:3d}}{\includegraphics[width=0.245\textwidth]{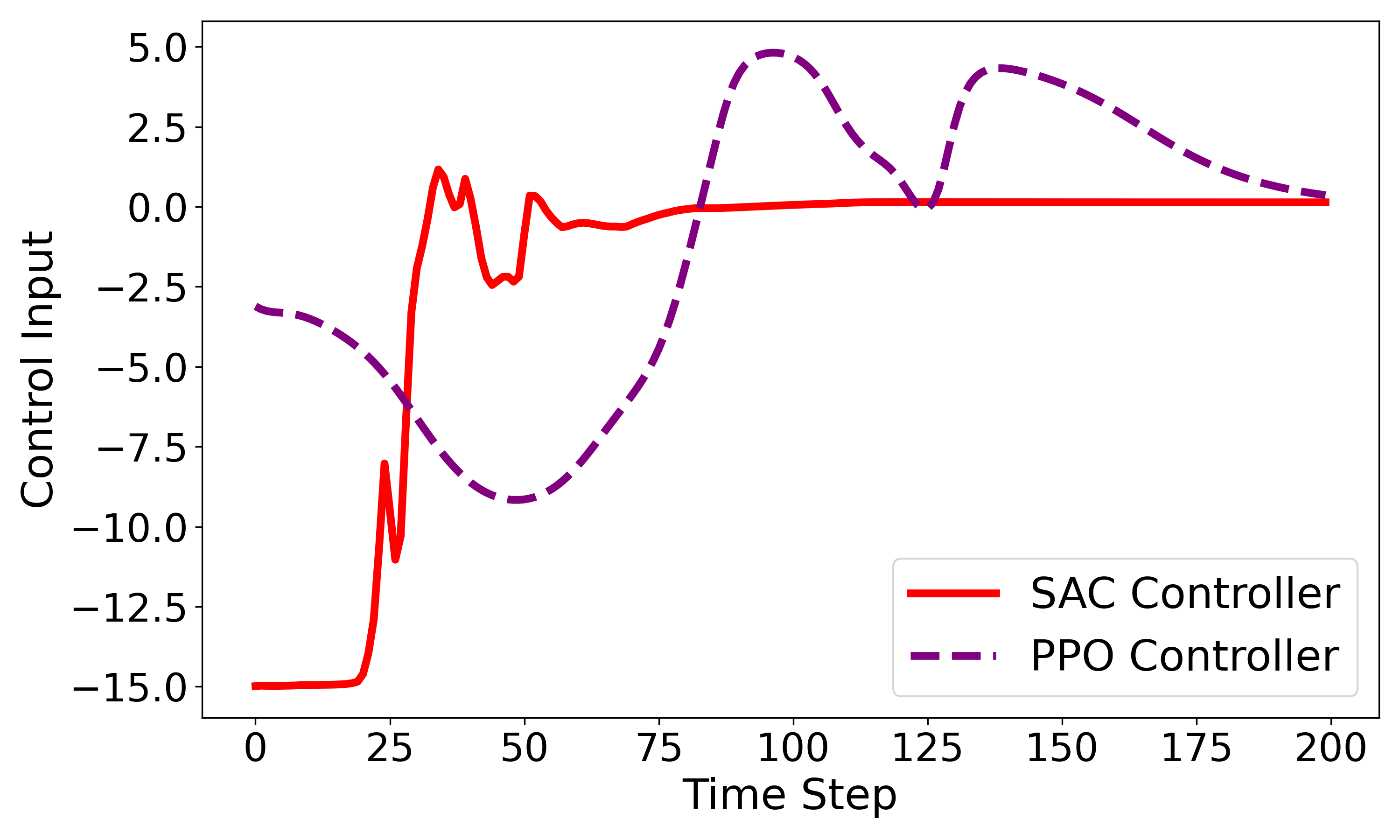}}\\
  \caption{Comparison of Lyapunov function values, control inputs, and RL value functions for the initial state $(\pi, 0)$. (a) Lyapunov function values over time for the baseline and DR controllers. (b) Control inputs over time for the baseline and DR controllers. (c) Value function values over time for the SAC and PPO algorithms. (d) Control inputs generated by the SAC and PPO algorithms.}
  \label{fig: compare_rl_1}
\end{figure*}

\begin{figure*}[ht]
  \centering
  \subcaptionbox{Lyapunov Function Values\label{fig:4a}}{\includegraphics[width=0.245\textwidth]{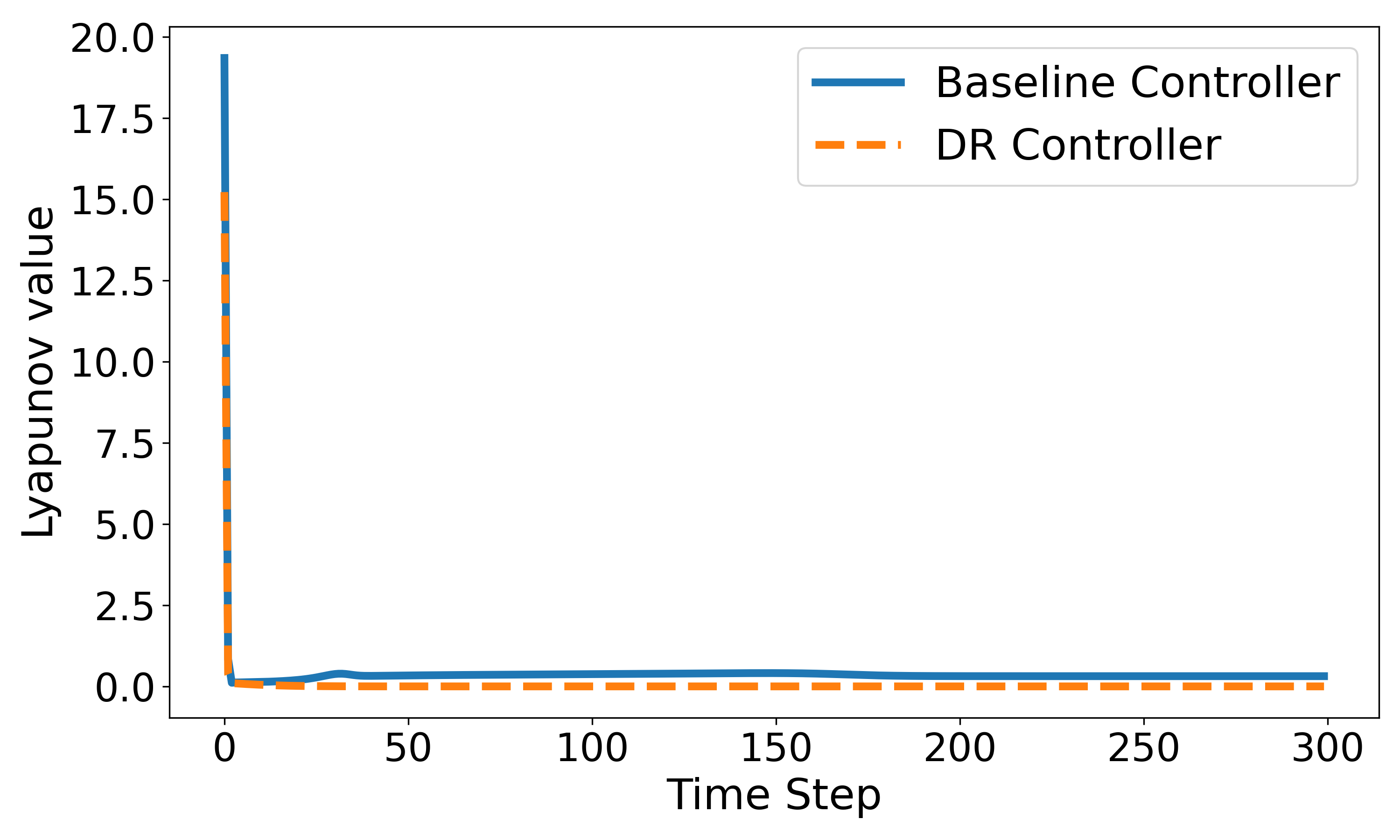}}%
  \hfill%
  \subcaptionbox{Control Inputs\label{fig:4b}}{\includegraphics[width=0.245\textwidth]{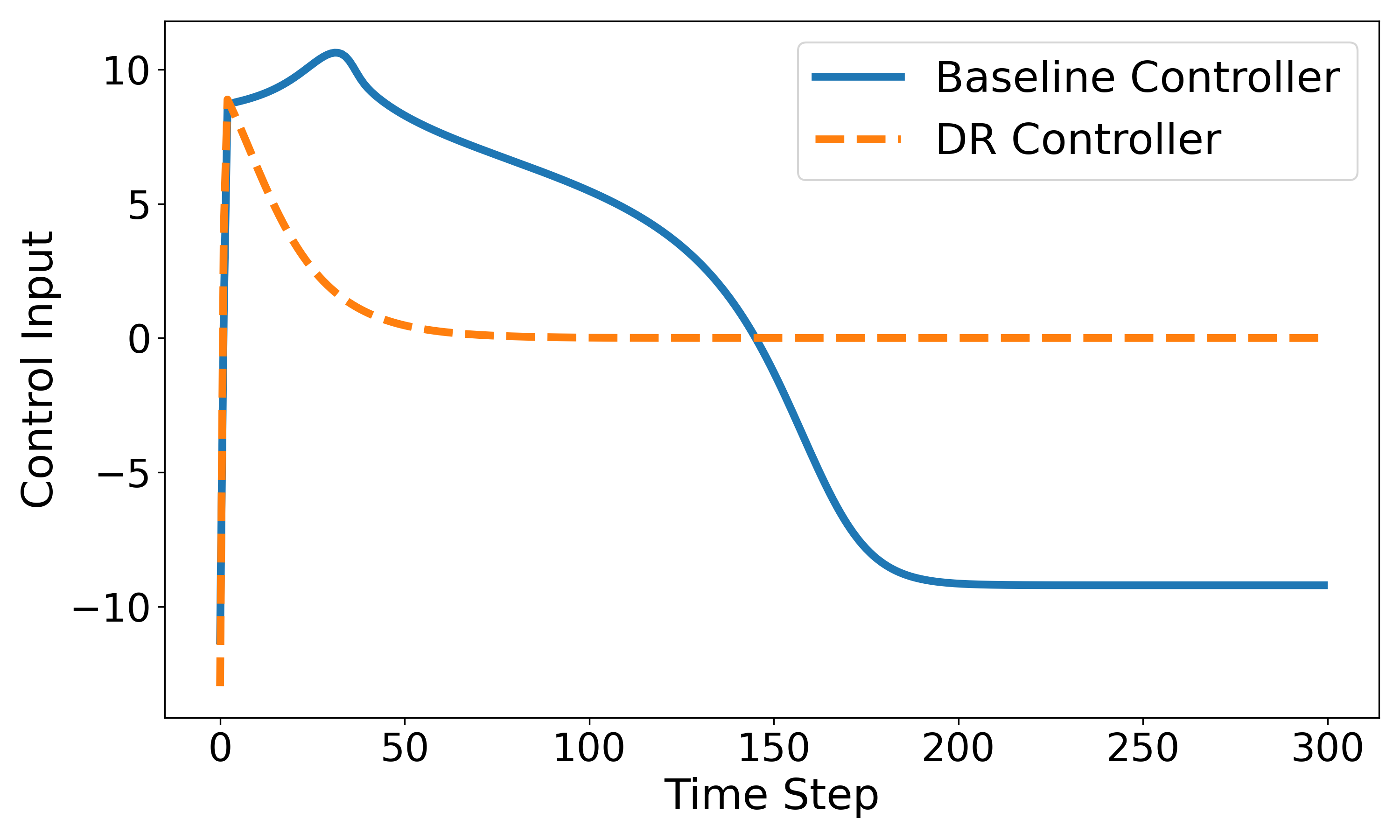}}
\subcaptionbox{RL Value Functions\label{fig:4c}}{\includegraphics[width=0.245\textwidth]{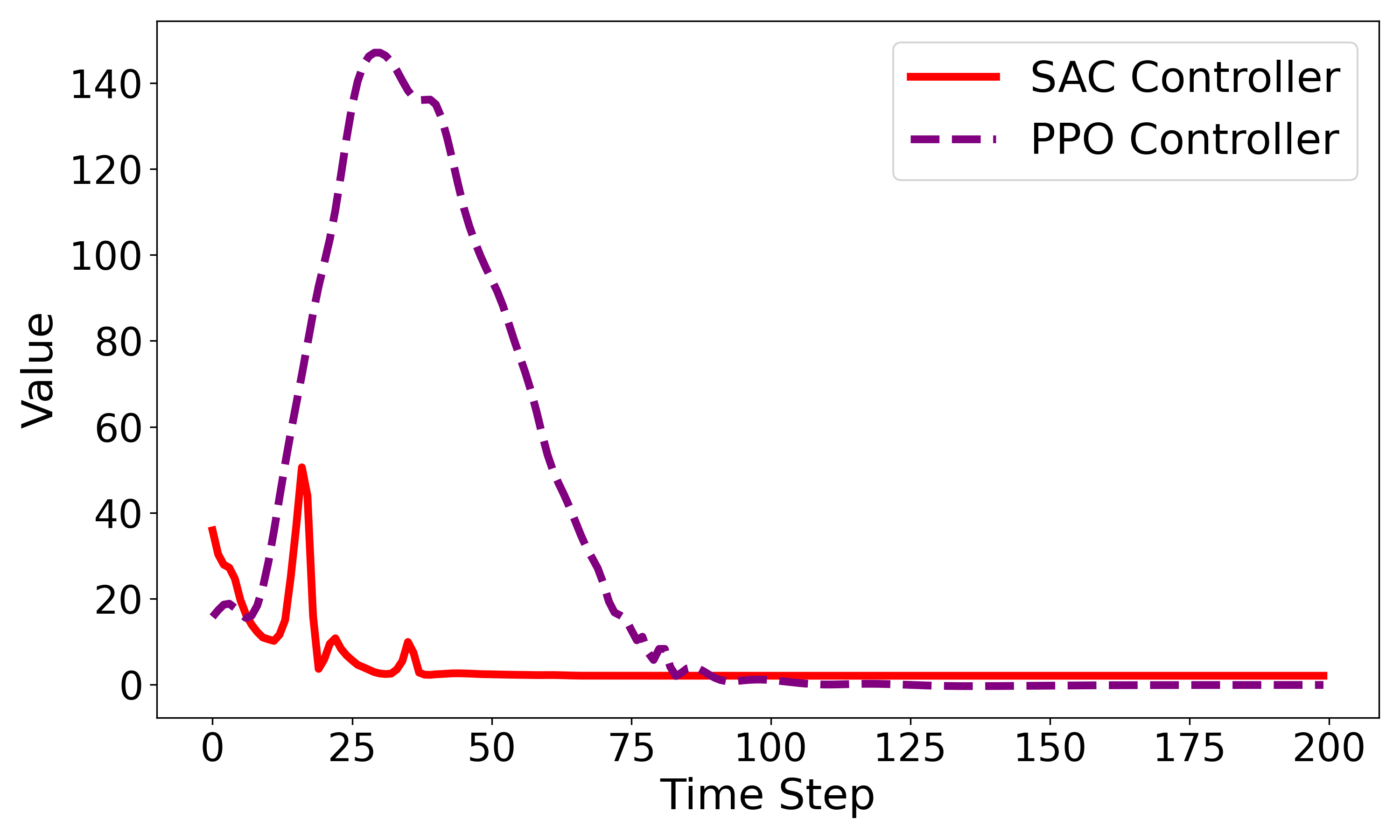}}%
  \hfill%
  \subcaptionbox{RL Control Inputs\label{fig:4d}}{\includegraphics[width=0.245\textwidth]{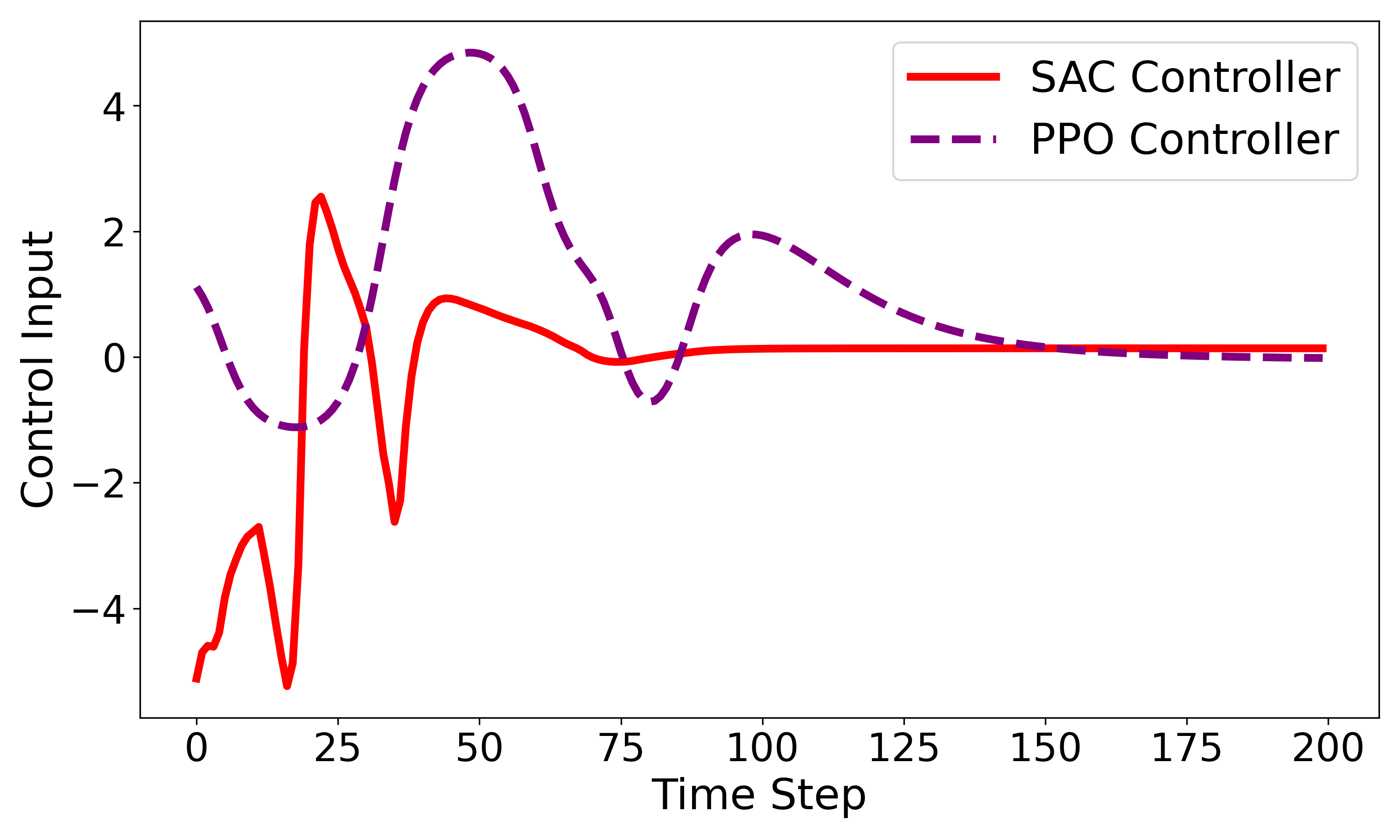}}\\
  \caption{Comparison of Lyapunov function values, control inputs, and RL value functions for the initial state $(-\pi/2, 5.5)$. (a) Lyapunov function values over time for the baseline and DR controllers. (b) Control inputs over time for the baseline and DR controllers. (c) Value function values over time for the SAC and PPO algorithms. (d) Control inputs generated by the SAC and PPO algorithms.}
  \label{fig: compare_rl_2}
\end{figure*}

The inverted pendulum is a standard nonlinear control problem for testing control methods. The system consists of two state variables,  angular position $\theta$ and angular velocity $\dot{\theta}$, and one control input $u$. The system dynamics are:
\begin{equation}
\label{eq: pendulum_dynamics}
    \underbrace{\left[ \begin{matrix}
    \dot{\theta} \\
    \ddot{\theta}
    \end{matrix} \right]}_{\dot{\bfx}}
	= \bff(\bfx, \bfu) := \left[ \begin{matrix}
    \dot{\theta} \\
    \frac{mgl \sin{\theta} - b \dot{\theta}}{ml^2}
    \end{matrix} \right] + \left[ \begin{matrix}
    0 \\
    \frac{1}{ml^2}
    \end{matrix} \right] u 
\end{equation}
where $g = 9.81$ m/s$^2$ is the gravity acceleration, $m = 1.0$ kg is the mass, $l = 1.0$ m is the length, and $b = 0.13$ N·m·s/rad is the damping coefficient. The equilibrium state is defined as $[\theta, \dot{\theta}] = [0,0]$, which corresponds to the upright equilibrium position.

For training both the baseline and the DR controllers, we generate a training dataset by uniformly sampling $\{ \bfx_i\}_{i \in [3600]}$ from the box region defined by $0 \leq \theta \leq 2\pi$ and $-8 \leq \dot{\theta} \leq 8$. 

We assume uncertainties in mass and damping, represented by \(\xi_1\) and \(\xi_2\), respectively. The uncertain system dynamics are:
\begin{equation}\label{eq: uncertain_pendulum_dynamics}
\dot{\bfx} = \left[ \begin{matrix}
\dot{\theta} \\
\frac{(m+\xi_1)gl \sin{\theta} - (b+\xi_2) \dot{\theta}}{(m+\xi_1)l^2}
\end{matrix} \right] + \left[ \begin{matrix}
0 \\
\frac{1}{(m+\xi_1)l^2}
\end{matrix} \right] u.
\end{equation}
Through first-order Taylor expansion around the nominal parameters, we derive the perturbation matrices $\bfW = [\bfw_1, \bfw_2]$ for the inverted pendulum subject to model uncertainty: 
\begin{equation}
\bfw_1(\bfx, \bfu) = \left[ \begin{matrix} 0 \\ \frac{b \dot{\theta} - u}{m^2 l^2} \end{matrix} \right], \quad
\bfw_2(\bfx, \bfu) = \left[ \begin{matrix} 0 \\ -\frac{\dot{\theta}}{ml^2} \end{matrix} \right] . 
\end{equation}

We assume that a set of offline samples $\{\bfxi_i\}_{i=1}^5$ of the uncertainties $\bfxi = [\xi_1, \xi_2]^\top$ are available for training the controller and certificate pairs. We take $\xi_1 \sim \mathcal{U}(-0.04, 0.08)$, $\xi_2 \sim \mathcal{N}(0.0, 0.02)$, where $\mathcal{U}$ and $\mathcal{N}$ denote uniform and normal distributions, respectively. However, during test time, the test uncertainty parameters are set to $\xi_1 = 0.1$ and $\xi_2 = 0.05$. These test values represent a significant deviation from the training distributions, allowing us to evaluate the controllers' robustness to distributional shifts in parameters.

To train the baseline controller, we compute the average mass $\tilde{m}$ and the average damping $\tilde{b}$ from the five offline samples. These average values are then used in the baseline system dynamics and the loss function~\eqref{eq: clf_loss} to learn the baseline controller and Lyapunov function pair. 

In contrast, for training the DR controller, we set the Wasserstein radius to $r = 0.01$ and the risk tolerance to $\epsilon = 0.1$. The $5$ uncertainty samples $\{\bfxi_i\}_{i=1}^5$ are then used in the loss function~\eqref{eq: dr_lf_loss} to learn the DR controller and its corresponding Lyapunov function. 


Fig.~\ref{fig: inverted_pendulum_traj} illustrates the performance of the baseline and DR controllers with the test uncertainty parameters $\xi_1 = 0.1$ and $\xi_2 = 0.05$. To assess the controllers' performance, we randomly sample 10 initial states within the box region defined by $0 \leq \theta \leq 2\pi$ and $-6 \leq \dot{\theta} \leq 6$. Each controller is applied to the system and the resulting trajectories are simulated.

In Fig.~\ref{fig:2a}, we observe that the baseline controller, trained using only the average mass and damping values, fails to stabilize the inverted pendulum system to the desired upright equilibrium. 
For all the $10$ initial states,  the baseline controller's trajectories converge to states that are not the desired equilibrium. This can be attributed to the increased damping ($\xi_2 = 0.05$) and the heavier pendulum mass ($\xi_1 = 0.1$) in the test scenario. The baseline controller, designed based on the average parameter values, lacks the necessary robustness to compensate for the increased damping and mass. 

On the other hand, Fig.~\ref{fig:2b} shows that all trajectories converge to the desired equilibrium under the DR controller. Despite the presence of distributional shift in the model uncertainty, the DR controller successfully stabilizes the inverted pendulum in an upright position. This robustness can be attributed to the DR controller's training process, which explicitly takes into account the distributional information of the uncertainty and optimizes for worst-case performance within the constructed ambiguity set.

To further investigate the performance of the proposed approach, we compare it with state-of-the-art reinforcement learning (RL) algorithms, such as the Soft Actor-Critic algorithm (SAC) \cite{haarnoja2018soft} and the Proximal Policy Optimization algorithm (PPO) \cite{schulman2017proximal}. The reward function used for training the RL algorithms is defined as $r = -(\theta^2 + 0.1 \dot{\theta}^2 + 0.001 u^2)$.

It is worth noting that our approach and the RL algorithms differ in their sample complexity. We use a dataset of $3600$ state samples uniformly sampled from the region of interest for training both the baseline and the DR controllers. In contrast, the stable baselines3 implementations \cite{stable-baselines3} of SAC and PPO require a significantly larger number of environment interactions, with a total of 2.048 billion steps for SAC and 4.096 billion steps for PPO. This higher sample complexity of the RL algorithms can be attributed to their model-free nature, as they learn directly from experience without explicitly using a dynamics model. Our approach incorporates the system dynamics model through Lyapunov theory, leading to a lower sample complexity.

Figs.~\ref{fig: compare_rl_1} and~\ref{fig: compare_rl_2} present a qualitative comparison of the baseline controller, our proposed DR controller, and the two RL policies over time for two different initial states. In Fig.~\ref{fig: compare_rl_1}, the initial state is the downright position $(\theta, \dot{\theta}) = (\pi, 0)$, from which the pendulum has to swing up to the desired equilibrium. We observe that the Lyapunov function corresponding to the baseline controller fails to converge to zero, while the proposed DR Lyapunov function values successfully converge to zero, as illustrated in Fig.~\ref{fig:3a}. Although the value functions learned by the two RL algorithms eventually reach zero, they exhibit non-monotonic behavior, indicating that the stability of the system cannot be certified using these learned value functions (Fig.~\ref{fig:3c}).

In Fig.~\ref{fig: compare_rl_2}, we change the initial state to $(-\pi/2, 5.5)$. In this scenario, the non-monotonic behavior of the value functions learned by the RL algorithms becomes more apparent, as shown in Fig.~\ref{fig:4c}. This observation highlights the advantage of the proposed DR approach, which provides provable stability guarantees through the use of a certified Lyapunov function. The DR controller ensures that the Lyapunov function values monotonically decrease towards zero with high probability (Fig.~\ref{fig:4a}), thereby enabling robustness to distributional shifts in the system parameters.

In conclusion, while the RL algorithms can learn control policies effectively, their lack of provable stability guarantees may limit their applicability in real-world control and robot systems. In contrast, the proposed DR approach offers a principled way to design controllers that are robust to distributional shifts and provide certifiable stability guarantees.

\begin{remark}[\textbf{Impact of Input Constraints on Controller Performance}]\label{rm: input_constraint_on_LF_learning}
{\rm In practical applications, control inputs are often subject to physical limitations, such as actuator saturation. These constraints can significantly impact the performance and stability of the controlled system. In our implementation, we enforce control input constraints using the $\text{clamp}(\cdot, \bfu_{\text{min}}, \bfu_{\text{max}})$ function, which ensures that the output remains within the specified control limits. Specifically, for the inverted pendulum system, we constrain the input $u$ within the bounds $-15 \leq u \leq 15$.

To investigate the effect of input constraints on our Lyapunov-based approach and on the reinforcement learning algorithms, we also considered a scenario where the control bounds are tightened to $-9 \leq u \leq 9$. Under these stricter input constraints, the Lyapunov-based approach failed to learn a controller with a valid Lyapunov function certificate. This failure occurs because the limited control authority hinders the ability of the Lyapunov-based controller to drive the system towards the desired equilibrium state, which is having the pendulum in the upright position. 

Interestingly, some reinforcement learning algorithms, such as PPO \cite{schulman2017proximal}, can still learn effective control policies even under tightened input constraints. These RL algorithms have the flexibility to discover alternative control strategies that exploit the system dynamics to achieve the desired goal. For instance, the RL controller may learn to swing the pendulum in one direction, leveraging gravity to gain acceleration, and then stabilize the pendulum in the upright position from the opposite direction. This swinging behavior allows the RL controller to overcome the limitations imposed by the reduced control bounds and successfully drive the system towards the equilibrium.
} \demo
\end{remark}


\subsection{Mountain Car}

\begin{figure}[t]
  \centering
  \subcaptionbox{Baseline Trajectories\label{fig:5a}}{\includegraphics[width=0.49\linewidth]{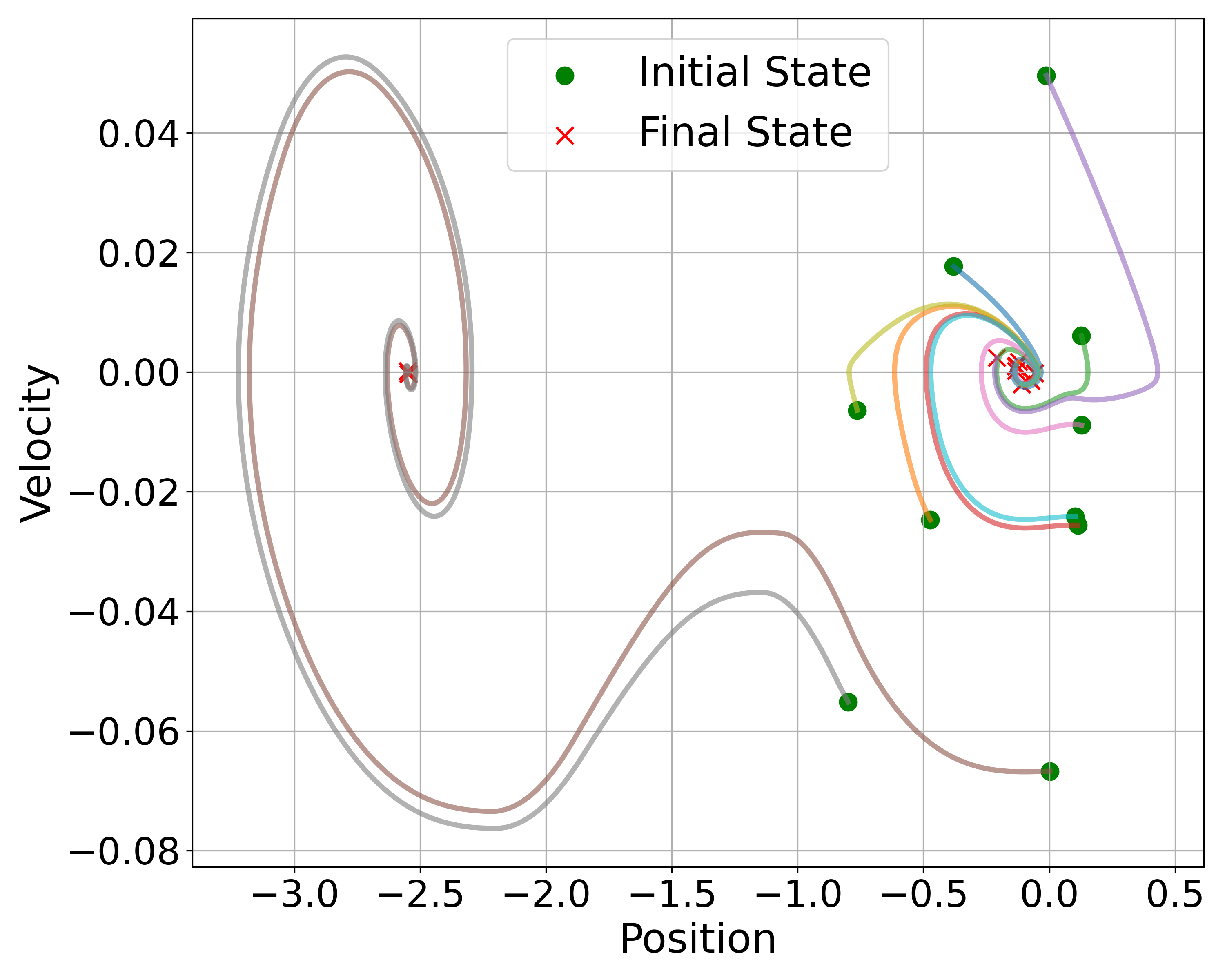}}%
  \hfill%
  \subcaptionbox{DR Trajectories\label{fig:5b}}{\includegraphics[width=0.49\linewidth]{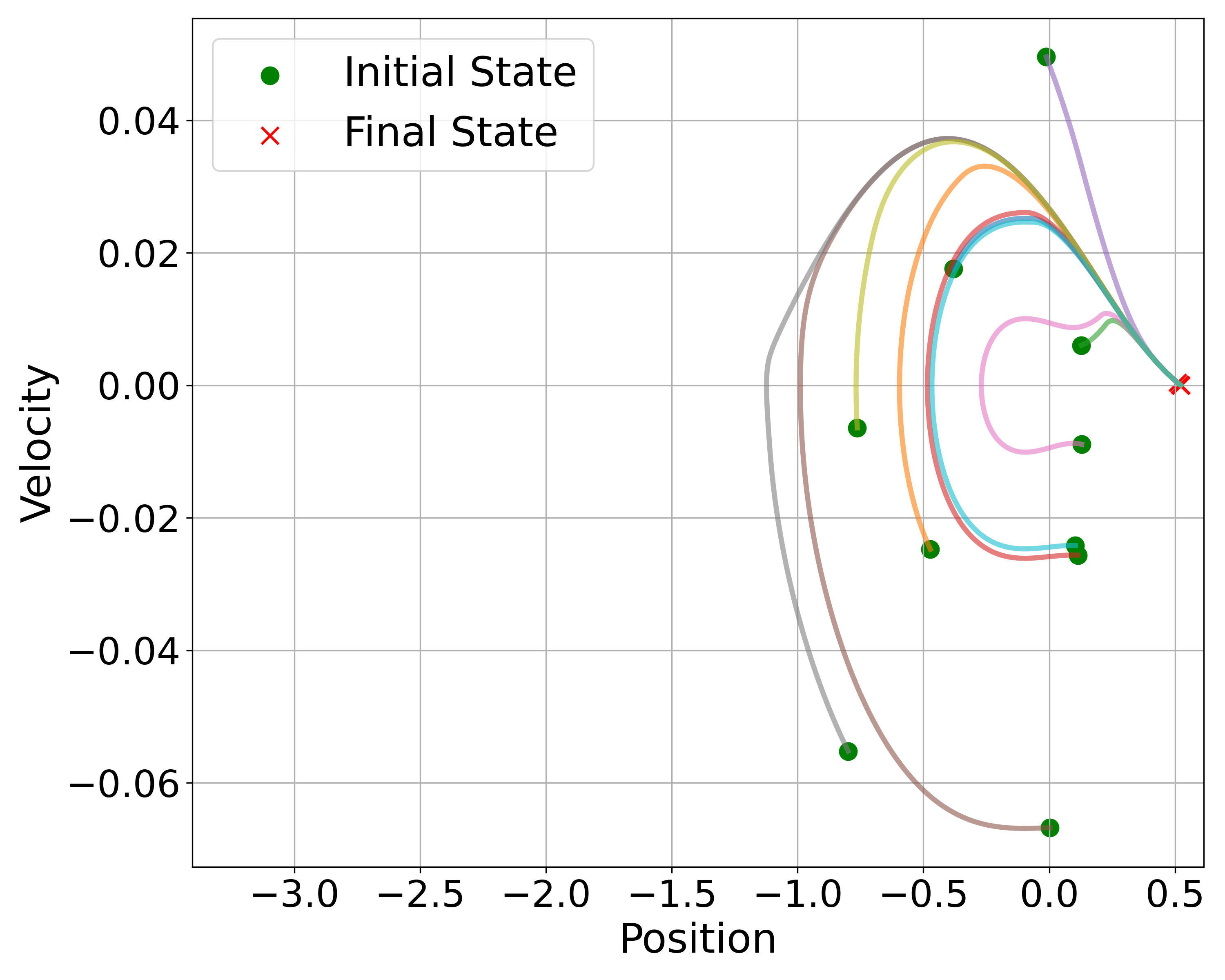}}\\
  \caption{Comparison of trajectories for the mountain car system with test case power parameter $p=0.0012$. The $10$ random sampled initial states are marked as green dots while the final states are marked as red crosses. In (a), the baseline controller, trained using the average power from offline observations, fails to stabilize the car to the desired equilibrium $(\pi/6, 0)$. Instead, (b) shows the distributionally robust (DR) controller successfully stabilizing the car to the top of the mountain for every initial state, demonstrating improved robustness to distributional shifts in system parameters.}
  \label{fig: moountain_car_traj}
\end{figure}

We consider a car placed at the bottom of a sinusoidal valley. The goal is to drive the car to the top of the mountain. The system has two states, velocity $v$ and position $x$ of the car, and one control input $u$, representing the directional force applied on the car. The dynamics model is given by~\cite{Moore90efficientmemory-based}:
\begin{equation}
\label{eq: mountain_cars}
    \underbrace{\left[ \begin{matrix}
    \dot{x} \\
    \dot{v}
    \end{matrix} \right]}_{\dot{\bfx}}
	= \bff(\bfx, \bfu) := \left[ \begin{matrix}
    \dot{v} \\
    -0.0025 \cos(3x)
    \end{matrix} \right] + \left[ \begin{matrix}
    0 \\
    p
    \end{matrix} \right] u 
\end{equation}
where $p = 0.0015$ denotes the constant power of the car. The desired equilibrium is $(x, v) = (\pi/6, 0)$, corresponding to the peak of the sinusoidal valley. 
For training both the baseline and the DR controllers, we generate a training dataset by uniformly sampling $\{ \bfx_i\}_{i \in [1600]}$ from the box region defined by $-2 \leq x \leq 2$ and $-0.4 \leq v \leq 0.4$. The control input is constrained within the bounds $-2 \leq u \leq 2$. We consider uncertainties in the car power~$p$, represented by \(\bfxi\). The uncertain system dynamics can be expressed as:
\begin{equation}\label{eq: uncertain_mountain_car}
\dot{\bfx} = \bff(\bfx, \bfu) + \left[ \begin{matrix} 0 \\ 
u
\end{matrix} \right] \bfxi.
\end{equation}
We take $\{\bfxi_i \}_{i=1}^3$ offline uncertainty samples from the normal distribution $\calN(0.0, 0.0002)$ are available. However, during test time, the uncertainty parameters are set to $\bfxi = -0.0003$.

For the baseline controller, we calculate the average power $\tilde{p}$ from the available power samples and incorporate it into the system dynamics~\eqref{eq: mountain_cars} during the learning process of the baseline Lyapunov-stable controller. In contrast, for training the DR controller, we set the Wasserstein radius to $r = 0.0001$ and the risk tolerance to $\epsilon = 0.1$. The offline samples $\{\bfxi_i \}_{i \in [3]}$ are then utilized in the loss function~\eqref{eq: dr_lf_loss} to learn the controller and Lyapunov function pair simultaneously.

To evaluate the performance of the baseline and DR controllers on the mountain car system, we consider a test scenario with uncertainty parameters set to $\bfxi = -0.0003$, representing a decrease in the car's power. We randomly initialize 10 starting states within the box region defined by $-1 \leq x \leq 1$ and $-0.1 \leq v \leq 0.1$. The controllers are then applied to the system, and the resulting trajectories are simulated to assess their effectiveness.

Fig.~\ref{fig:5a} reveals that the baseline controller, which is trained on the average power values, fails to drive the car to the desired equilibrium at the top of the mountain. The trajectories corresponding to all 10 initial states fail to converge to the desired goal. The baseline controller's design, based on average parameter values, lacks the robustness necessary to adapt to the decreased power.

In contrast, Fig.~\ref{fig:5b} showcases the superior performance of the DR controller. Despite the presence of distributional shift in the system dynamics, the DR controller successfully drives all trajectories to the desired equilibrium at the mountain's peak. By explicitly considering the potential variations in the car's power during training, the DR controller is able to adapt and maintain its effectiveness for out-of-distribution parameter uncertainties.

\section{Conclusions}
\label{sec: conclusion}

In this article, we presented an approach for the joint synthesis of distributionally robust controllers and Lyapunov certificates suitable for nonlinear systems with model uncertainty. Our key contributions lie in the development of a distributionally robust Lyapunov stability formulation, leveraging finite samples of the model uncertainty parameters to address the challenge of stabilizing the uncertain system with guarantees. Through both theoretical analysis and experimental validation, we demonstrated that the system governed by the learned controller exhibits asymptotic stability with high probability, even in the presence of out-of-distribution model uncertainties.

Our approach contributes a new perspective to controlling systems with model uncertainty, offering a promising direction for certificate-based reinforcement learning in real systems. However, scalability challenges arise from the need to satisfy the Lyapunov condition for uniformly sampled data in the state space, leading to an exponential growth in the number of training samples as the dimension of the system state increases. Future research will focus on adapting our methods to more complex robotic systems. This includes exploring more efficient learning algorithms, considering relaxed stability notions, and establishing connections between Lyapunov theory and model-free reinforcement learning.

%
%
\bibliographystyle{ieeetr}
\bibliography{ref}

\begin{thebibliography}{10}

\bibitem{long2023dro_lf}
K.~Long, Y.~Yi, J.~Cortes, and N.~Atanasov, ``Distributionally robust {L}yapunov function search under uncertainty,'' in {\em Learning for Dynamics and Control Conference}, pp.~864--877, PMLR, 2023.

\bibitem{slotine1991applied}
J.-J.~E. Slotine, W.~Li, {\em et~al.}, {\em Applied nonlinear control}, vol.~199.
\newblock Prentice hall Englewood Cliffs, NJ, 1991.

\bibitem{Artstein1983StabilizationWR}
Z.~Artstein, ``Stabilization with relaxed controls,'' {\em Nonlinear Analysis-theory Methods \& Applications}, vol.~7, pp.~1163--1173, 1983.

\bibitem{SONTAG1989117}
E.~Sontag, ``{A ‘universal’ construction of Artstein's theorem on nonlinear stabilization},'' {\em Systems \& Control Letters}, vol.~13, no.~2, pp.~117--123, 1989.

\bibitem{haddad2008nonlinear}
W.~M. Haddad and V.~Chellaboina, {\em Nonlinear dynamical systems and control: a {L}yapunov-based approach}.
\newblock Princeton university press, 2008.

\bibitem{parrilo2000structured}
P.~Parrilo, {\em Structured semidefinite programs and semialgebraic geometry methods in robustness and optimization}.
\newblock California Institute of Technology, 2000.

\bibitem{Papachristo_2002_sos_lf}
A.~Papachristodoulou and S.~Prajna, ``On the construction of {L}yapunov functions using the sum of squares decomposition,'' in {\em Proceedings of the 41st IEEE Conference on Decision and Control, 2002.}, vol.~3, pp.~3482--3487, 2002.

\bibitem{jarvis_clf_sos}
Z.~Jarvis-Wloszek, R.~Feeley, W.~Tan, K.~Sun, and A.~Packard, ``Some controls applications of sum of squares programming,'' in {\em 42nd IEEE International Conference on Decision and Control}, vol.~5, pp.~4676--4681, 2003.

\bibitem{Chang2019NeuralLC}
Y.-C. Chang, N.~Roohi, and S.~Gao, ``Neural {L}yapunov control,'' in {\em Advances in Neural Information Processing Systems}, vol.~32, 2019.

\bibitem{Richards2018TheLN}
S.~M. Richards, F.~Berkenkamp, and A.~Krause, ``The {L}yapunov neural network: Adaptive stability certification for safe learning of dynamical systems,'' in {\em Conference on Robot Learning}, pp.~466--476, PMLR, 2018.

\bibitem{dai_2021_lyapunov}
H.~Dai, B.~Landry, L.~Yang, M.~Pavone, and R.~Tedrake, ``{L}yapunov-stable neural-network control,'' in {\em Proceedings of Robotics: Science and Systems}, (Virtual), July 2021.

\bibitem{gaby2021lyapunov_net}
N.~Gaby, F.~Zhang, and X.~Ye, ``{L}yapunov-net: A deep neural network architecture for {L}yapunov function approximation,'' in {\em 2022 IEEE 61st Conference on Decision and Control (CDC)}, pp.~2091--2096, 2022.

\bibitem{li2023task}
T.~Li and N.~Figueroa, ``Task generalization with stability guarantees via elastic dynamical system motion policies,'' in {\em Conference on Robot Learning}, pp.~3485--3517, PMLR, 2023.

\bibitem{wu2023neural}
J.~Wu, A.~Clark, Y.~Kantaros, and Y.~Vorobeychik, ``Neural {L}yapunov control for discrete-time systems,'' {\em Advances in neural information processing systems}, vol.~36, pp.~2939--2955, 2023.

\bibitem{zhang2024learning}
S.~Zhang and C.~Fan, ``Learning to stabilize high-dimensional unknown systems using {L}yapunov-guided exploration,'' in {\em 6th Annual Learning for Dynamics \& Control Conference}, pp.~52--67, PMLR, 2024.

\bibitem{Taylor_2019iros}
A.~Taylor, V.~Dorobantu, H.~Le, Y.~Yue, and A.~Ames, ``Episodic learning with control {L}yapunov functions for uncertain robotic systems,'' in {\em 2019 IEEE/RSJ International Conference on Intelligent Robots and Systems (IROS)}, pp.~6878--6884, 2019.

\bibitem{Castaneda_GPCLF_ACC21}
F.~Castañeda, J.~Choi, B.~Zhang, C.~Tomlin, and K.~Sreenath, ``Gaussian process-based min-norm stabilizing controller for control-affine systems with uncertain input effects and dynamics,'' in {\em 2021 American Control Conference (ACC)}, pp.~3683--3690, 2021.

\bibitem{Long2022RAL}
K.~Long, V.~Dhiman, M.~Leok, J.~Cortés, and N.~Atanasov, ``Safe control synthesis with uncertain dynamics and constraints,'' {\em IEEE Robotics and Automation Letters}, vol.~7, no.~3, pp.~7295--7302, 2022.

\bibitem{PM-JC:23-auto}
P.~Mestres and J.~Cort\'es, ``Feasibility and regularity analysis of safe stabilizing controllers under uncertainty,'' {\em Automatica}, 2023.
\newblock Submitted.

\bibitem{PM-KL-NA-JC:23-csl}
P.~Mestres, K.~Long, N.~Atanasov, and J.~Cortés, ``Feasibility analysis and regularity characterization of distributionally robust safe stabilizing controllers,'' {\em IEEE Control Systems Letters}, vol.~8, pp.~91--96, 2024.

\bibitem{Long_learningcbf_ral21}
K.~Long, C.~Qian, J.~Cortés, and N.~Atanasov, ``Learning barrier functions with memory for robust safe navigation,'' {\em IEEE Robotics and Automation Letters}, vol.~6, no.~3, pp.~4931--4938, 2021.

\bibitem{AS-DD-AR:21}
A.~Shapiro, D.~Dentcheva, and A.~Ruszczyński, {\em Lectures on Stochastic Programming}.
\newblock Society for Industrial and Applied Mathematics, 2009.

\bibitem{Esfahani2018DatadrivenDR}
P.~M. Esfahani and D.~Kuhn, ``Data-driven distributionally robust optimization using the {W}asserstein metric: performance guarantees and tractable reformulations,'' {\em Mathematical Programming}, vol.~171, pp.~115--166, 2018.

\bibitem{Hota2019DataDrivenCC}
A.~R. Hota, A.~K. Cherukuri, and J.~Lygeros, ``Data-driven chance constrained optimization under {W}asserstein ambiguity sets,'' in {\em 2019 American Control Conference (ACC)}, pp.~1501--1506, 2019.

\bibitem{yang2020wasserstein}
I.~Yang, ``Wasserstein distributionally robust stochastic control: A data-driven approach,'' {\em IEEE Transactions on Automatic Control}, vol.~66, no.~8, pp.~3863--3870, 2020.

\bibitem{coulson2021distributionally}
J.~Coulson, J.~Lygeros, and F.~D{\"o}rfler, ``Distributionally robust chance constrained data-enabled predictive control,'' {\em IEEE Transactions on Automatic Control}, vol.~67, no.~7, pp.~3289--3304, 2021.

\bibitem{bahari_2022_safedro_rl}
A.~B. Kordabad, R.~Wisniewski, and S.~Gros, ``Safe reinforcement learning using {W}asserstein distributionally robust {MPC} and chance constraint,'' {\em IEEE Access}, vol.~10, pp.~130058--130067, 2022.

\bibitem{lathrop2021distributionally}
P.~Lathrop, B.~Boardman, and S.~Mart{\'\i}nez, ``Distributionally safe path planning: {W}asserstein safe {RRT},'' {\em IEEE Robotics and Automation Letters}, vol.~7, no.~1, pp.~430--437, 2021.

\bibitem{Hakobyan_dr_ddp}
A.~Hakobyan and I.~Yang, ``Distributionally robust differential dynamic programming with {W}asserstein distance,'' {\em IEEE Control Systems Letters}, vol.~7, pp.~2329--2334, 2023.

\bibitem{li2021distributionally}
B.~Li, Y.~Tan, A.-G. Wu, and G.-R. Duan, ``A distributionally robust optimization based method for stochastic model predictive control,'' {\em IEEE Transactions on Automatic Control}, vol.~67, no.~11, pp.~5762--5776, 2021.

\bibitem{aolaritei2023wasserstein}
L.~Aolaritei, M.~Fochesato, J.~Lygeros, and F.~D{\"o}rfler, ``Wasserstein tube {MPC} with exact uncertainty propagation,'' in {\em 2023 62nd IEEE Conference on Decision and Control (CDC)}, pp.~2036--2041, IEEE, 2023.

\bibitem{li2023_drc_output}
B.~Li, T.~Guan, L.~Dai, and G.-R. Duan, ``Distributionally robust model predictive control with output feedback,'' {\em IEEE Transactions on Automatic Control}, vol.~69, no.~5, pp.~3270--3277, 2024.

\bibitem{micheli2022data}
F.~Micheli, T.~Summers, and J.~Lygeros, ``Data-driven distributionally robust mpc for systems with uncertain dynamics,'' in {\em 2022 IEEE 61st Conference on Decision and Control (CDC)}, pp.~4788--4793, IEEE, 2022.

\bibitem{taskesen2024distributionally}
B.~Taskesen, D.~Iancu, {\c{C}}.~Ko{\c{c}}yi{\u{g}}it, and D.~Kuhn, ``Distributionally robust linear quadratic control,'' {\em Advances in Neural Information Processing Systems}, vol.~36, 2024.

\bibitem{Boyd_LMI_control}
S.~Boyd, L.~El~Ghaoui, E.~Feron, and V.~Balakrishnan, {\em Linear Matrix Inequalities in System and Control Theory}.
\newblock Society for Industrial and Applied Mathematics, 1994.

\bibitem{HilbertUeberDD}
D.~Hilbert, ``Ueber die darstellung definiter formen als summe von formenquadraten,'' {\em Mathematische Annalen}, vol.~32, pp.~342--350, September 1888.

\bibitem{boffi2021learning}
N.~Boffi, S.~Tu, N.~Matni, J.-J. Slotine, and V.~Sindhwani, ``Learning stability certificates from data,'' in {\em Conference on Robot Learning}, pp.~1341--1350, PMLR, 2021.

\bibitem{zhou2022neural}
R.~Zhou, T.~Quartz, H.~De~Sterck, and J.~Liu, ``Neural {L}yapunov control of unknown nonlinear systems with stability guarantees,'' in {\em Advances in Neural Information Processing Systems}, vol.~35, 2022.

\bibitem{dawson_2022_robust_CLBF}
C.~Dawson, Z.~Qin, S.~Gao, and C.~Fan, ``Safe nonlinear control using robust neural {L}yapunov-barrier functions,'' in {\em Conference on Robot Learning}, vol.~164, pp.~1724--1735, PMLR, 2022.

\bibitem{Dawson_2022_survey}
C.~Dawson, S.~Gao, and C.~Fan, ``Safe control with learned certificates: A survey of neural {L}yapunov, barrier, and contraction methods for robotics and control,'' {\em IEEE Transactions on Robotics}, vol.~39, no.~3, pp.~1749--1767, 2023.

\bibitem{westenbroek2022lyapunov}
T.~Westenbroek, F.~Castaneda, A.~Agrawal, S.~Sastry, and K.~Sreenath, ``Lyapunov design for robust and efficient robotic reinforcement learning,'' {\em arXiv preprint arXiv:2208.06721}, 2022.

\bibitem{lopez2024decomposing}
A.~Lopez and D.~Fridovich-Keil, ``Decomposing control lyapunov functions for efficient reinforcement learning,'' {\em arXiv preprint arXiv:2403.12210}, 2024.

\bibitem{Parys2015monent}
B.~P.~G. Van~Parys, D.~Kuhn, P.~J. Goulart, and M.~Morari, ``Distributionally robust control of constrained stochastic systems,'' {\em IEEE Transactions on Automatic Control}, vol.~61, no.~2, pp.~430--442, 2016.

\bibitem{Jiang2016DatadrivenCC}
R.~Jiang and Y.~Guan, ``Data-driven chance constrained stochastic program,'' {\em Mathematical Programming}, vol.~158, pp.~291--327, 2016.

\bibitem{Xie2021OnDR}
W.~Xie, ``On distributionally robust chance constrained programs with {W}asserstein distance,'' {\em Math. Program.}, vol.~186, pp.~115--155, 2021.

\bibitem{sagawa2019distributionally}
S.~Sagawa*, P.~W. Koh*, T.~B. Hashimoto, and P.~Liang, ``Distributionally robust neural networks,'' in {\em International Conference on Learning Representations}, 2020.

\bibitem{levine2020offline}
S.~Levine, A.~Kumar, G.~Tucker, and J.~Fu, ``Offline reinforcement learning: Tutorial, review, and perspectives on open problems,'' {\em arXiv preprint arXiv:2005.01643}, 2020.

\bibitem{FB-DB-JC-SM-DMT:21}
F.~Boso, D.~Boskos, J.~Cort\'es, S.~Mart{\'\i}nez, and D.~M. Tartakovsky, ``Dynamics of data-driven ambiguity sets for hyperbolic conservation laws with uncertain inputs,'' {\em SIAM Journal on Scientific Computing}, vol.~43, no.~3, pp.~A2102--A2129, 2021.

\bibitem{DB-JC-SM:21-tac}
D.~Boskos, J.~Cort\'es, and S.~Martinez, ``Data-driven ambiguity sets with probabilistic guarantees for dynamic processes,'' {\em IEEE Transactions on Automatic Control}, vol.~66, no.~7, pp.~2991--3006, 2021.

\bibitem{ashish_2023_dro}
A.~Cherukuri and A.~R. Hota, ``Consistency of distributionally robust risk- and chance-constrained optimization under {W}asserstein ambiguity sets,'' {\em IEEE Control Systems Letters}, vol.~5, no.~5, pp.~1729--1734, 2021.

\bibitem{DB-JC-SM:24-tac}
D.~Boskos, J.~Cort{\'e}s, and S.~Mart{\'i}nez, ``High-confidence data-driven ambiguity sets for time-varying linear systems,'' {\em IEEE Transactions on Automatic Control}, vol.~69, no.~2, pp.~797--812, 2024.

\bibitem{ren2022distributionally_ral}
A.~Z. Ren and A.~Majumdar, ``Distributionally robust policy learning via adversarial environment generation,'' {\em IEEE Robotics and Automation Letters}, vol.~7, no.~2, pp.~1379--1386, 2022.

\bibitem{long2023_clf_cbf_drccp}
K.~Long, Y.~Yi, J.~Cortés, and N.~Atanasov, ``Safe and stable control synthesis for uncertain system models via distributionally robust optimization,'' in {\em 2023 American Control Conference (ACC)}, pp.~4651--4658, 2023.

\bibitem{summers2018_dr_rrt}
T.~Summers, ``Distributionally robust sampling-based motion planning under uncertainty,'' in {\em 2018 IEEE/RSJ International Conference on Intelligent Robots and Systems (IROS)}, pp.~6518--6523, 2018.

\bibitem{sastry2013nonlinear}
S.~Sastry, {\em Nonlinear systems: analysis, stability, and control}, vol.~10.
\newblock Springer Science \& Business Media, 2013.

\bibitem{freeman_robust}
R.~Freeman and P.~Kototovic, {\em Robust Nonlinear Control Design}.
\newblock Cambridge, MA, USA: Birkh\"{a}user Boston Inc., 1996.

\bibitem{Rockafellar00optimizationof}
R.~T. Rockafellar and S.~Uryasev, ``Optimization of conditional value-at-risk,'' {\em Journal of Risk}, vol.~2, pp.~21--41, 2000.

\bibitem{Nemirovski2006ConvexAO}
A.~Nemirovski and A.~Shapiro, ``Convex approximations of chance constrained programs,'' {\em SIAM J. Optim.}, vol.~17, pp.~969--996, 2006.

\bibitem{kushner1967stochastic}
H.~J. Kushner and Kushner, {\em Stochastic stability and control}, vol.~33.
\newblock Academic press New York, 1967.

\bibitem{Teel_stochastic_stability}
A.~R. Teel, J.~P. Hespanha, and A.~Subbaraman, ``A converse {L}yapunov theorem and robustness for asymptotic stability in probability,'' {\em IEEE Transactions on Automatic Control}, vol.~59, no.~9, pp.~2426--2441, 2014.

\bibitem{culbertson2023input}
P.~Culbertson, R.~K. Cosner, M.~Tucker, and A.~D. Ames, ``Input-to-state stability in probability,'' in {\em 2023 62nd IEEE Conference on Decision and Control (CDC)}, pp.~5796--5803, IEEE, 2023.

\bibitem{steinhardt2012finite}
J.~Steinhardt and R.~Tedrake, ``Finite-time regional verification of stochastic non-linear systems,'' {\em The International Journal of Robotics Research}, vol.~31, no.~7, pp.~901--923, 2012.

\bibitem{santoyo2021barrier}
C.~Santoyo, M.~Dutreix, and S.~Coogan, ``A barrier function approach to finite-time stochastic system verification and control,'' {\em Automatica}, vol.~125, p.~109439, 2021.

\bibitem{weaver2018lipschitz}
N.~Weaver, {\em Lipschitz algebras}.
\newblock World Scientific, 2018.

\bibitem{fazlyab2019efficient}
M.~Fazlyab, A.~Robey, H.~Hassani, M.~Morari, and G.~Pappas, ``Efficient and accurate estimation of lipschitz constants for deep neural networks,'' in {\em Advances in Neural Information Processing Systems}, vol.~32, 2019.

\bibitem{hornik1989multilayer}
K.~Hornik, M.~Stinchcombe, and H.~White, ``Multilayer feedforward networks are universal approximators,'' {\em Neural networks}, vol.~2, no.~5, pp.~359--366, 1989.

\bibitem{leshno1993multilayer}
M.~Leshno, V.~Y. Lin, A.~Pinkus, and S.~Schocken, ``Multilayer feedforward networks with a nonpolynomial activation function can approximate any function,'' {\em Neural networks}, vol.~6, no.~6, pp.~861--867, 1993.

\bibitem{yang2024lyapunovstable}
L.~Yang, H.~Dai, Z.~Shi, C.-J. Hsieh, R.~Tedrake, and H.~Zhang, ``Lyapunov-stable neural control for state and output feedback: A novel formulation,'' in {\em Forty-first International Conference on Machine Learning}, 2024.

\bibitem{wang2021beta}
S.~Wang, H.~Zhang, K.~Xu, X.~Lin, S.~Jana, C.-J. Hsieh, and J.~Z. Kolter, ``{Beta-CROWN}: Efficient bound propagation with per-neuron split constraints for complete and incomplete neural network verification,'' {\em Advances in Neural Information Processing Systems}, vol.~34, 2021.

\bibitem{wu2024marabou}
H.~Wu, O.~Isac, A.~Zelji{\'c}, T.~Tagomori, M.~Daggitt, W.~Kokke, I.~Refaeli, G.~Amir, K.~Julian, S.~Bassan, {\em et~al.}, ``Marabou 2.0: A versatile formal analyzer of neural networks,'' {\em arXiv preprint arXiv:2401.14461}, 2024.

\bibitem{towers_gymnasium_2023}
M.~Towers, J.~K. Terry, A.~Kwiatkowski, J.~U. Balis, G.~d. Cola, T.~Deleu, M.~Goulão, A.~Kallinteris, A.~KG, M.~Krimmel, R.~Perez-Vicente, A.~Pierré, S.~Schulhoff, J.~J. Tai, A.~T.~J. Shen, and O.~G. Younis, ``Gymnasium,'' Mar. 2023.

\bibitem{paszke2019pytorch}
A.~Paszke, S.~Gross, F.~Massa, A.~Lerer, J.~Bradbury, G.~Chanan, T.~Killeen, Z.~Lin, N.~Gimelshein, L.~Antiga, {\em et~al.}, ``Pytorch: An imperative style, high-performance deep learning library,'' in {\em Advances in neural information processing systems}, vol.~32, 2019.

\bibitem{Adam}
D.~P. Kingma and J.~Ba, ``Adam: A method for stochastic optimization,'' {\em arXiv preprint arXiv:1412.6980}, 2015.

\bibitem{haarnoja2018soft}
T.~Haarnoja, A.~Zhou, P.~Abbeel, and S.~Levine, ``Soft actor-critic: Off-policy maximum entropy deep reinforcement learning with a stochastic actor,'' in {\em International conference on machine learning}, pp.~1861--1870, PMLR, 2018.

\bibitem{schulman2017proximal}
J.~Schulman, F.~Wolski, P.~Dhariwal, A.~Radford, and O.~Klimov, ``Proximal policy optimization algorithms,'' {\em arXiv preprint arXiv:1707.06347}, 2017.

\bibitem{stable-baselines3}
A.~Raffin, A.~Hill, A.~Gleave, A.~Kanervisto, M.~Ernestus, and N.~Dormann, ``Stable-baselines3: Reliable reinforcement learning implementations,'' {\em Journal of Machine Learning Research}, vol.~22, no.~268, pp.~1--8, 2021.

\bibitem{Moore90efficientmemory-based}
A.~W. Moore, ``Efficient memory-based learning for robot control,'' tech. rep., University of Cambridge, 1990.

\end{thebibliography}

\appendices

\end{document}